\documentclass[useAMS,usenatbib]{aa}  

\usepackage{txfonts}
\usepackage{graphicx}
\usepackage{amsmath}
\usepackage{tabularx}
\usepackage{inputenc}
\usepackage{supertabular}
\usepackage{ltxtable}
\usepackage{subfig}
\begin{document}

   \title{HADES RV Programme with HARPS-N at TNG.}

   \subtitle{VII. Rotation and activity of M-Dwarfs from time-series high-resolution  spectroscopy of chromospheric indicators \thanks{Based on: observations made with the Italian Telescopio Nazionale Galileo (TNG), operated on the island of La Palma by the INAF - Fundación Galileo Galilei at the Roche de Los Muchachos Observatory of the Instituto  de  Astrofísica  de  Canarias (IAC);  observations  made  with the  HARPS instrument  on  theESO 3.6-m telescope at La Silla Observatory (Chile)}}

      \author{A. Su\'{a}rez Mascare\~{n}o \inst{1,2} \and
          R. Rebolo \inst{1,3,4} \and
          J.~I. Gonz\'alez Hern\'andez \inst{1,3}  \and
          B. Toledo-Padrón\inst{1,3} \and
          M. Perger \inst{5} \and
          I. Ribas \inst{5} \and
          L. Affer\inst{6} \and
          G. Micela\inst{6} \and
          M. Damasso \inst{7} \and
          J. Maldonado \inst{6} \and
          E. González-Alvarez \inst{6,8} \and
          G. Leto \inst{9} \and
          I. Pagano \inst{9} \and
          G. Scandariato \inst{9} \and
          A. Sozzetti \inst{7} \and
          A. ~F. Lanza \inst{9} \and
          L. Malavolta \inst{10,11} \and
          R. Claudi \inst{10} \and
          R. Cosentino \inst{9,12} \and
          S. Desidera \inst{10} \and
          P. Giacobbe \inst{7} \and
          A. Maggio \inst{6} \and
          M. Rainer \inst{13} \and
          M. Esposito \inst{14} \and
          S. Benatti \inst{10} \and
          M. Pedani \inst{12} \and
          J. ~C. Morales \inst{5} \and 
          E. Herrero \inst{5} \and
          M. Lafarga \inst{5} \and 
          A. Rosich \inst{5} \and
          M. Pinamonti \inst{7}
      } 

          \institute{Instituto de Astrof\'{i}sica de Canarias, E-38205 La Laguna, Tenerife, Spain\\
\email{asm\_ext@iac.es}            \and                      
                                Observatoire Astronomique de l'Université de Genève, 1290 Versoix, Switzerland \\
\email{Alejandro.SuarezMascareno@unige.ch}          \and
                                Universidad de La Laguna, Dpto. Astrof\'{i}sica, E-38206 La Laguna, Tenerife, Spain \and
Consejo Superior de Investigaciones Cient{\'\i}ficas, 28006, Madrid, Spain  \and
                                Institut de Ciènces de l'Espai (CSIC-IEEC), Campus UAB, Carrer de Can Magrans s/n, 08193 Cerdanyola del Vallés, Spain \and
                                INAF - Osservatorio Astronomico di Palermo, Piazza del Parlamento 1, 90134 Palermo, Italy \and
                    INAF - Osservatorio Astrofisico di Torino, via Osservatorio 20, 10025 Pino Torinese, Italy \and
                    Dipartimento di Fisica e Chimica, Universit\`{a} di Palermo, Piazza del Parlamento 1, 90134 Palermo, Italy\and
                    INAF - Osservatorio Astrofisico di Catania, via S. Sofia 78, 95123 Catania, Italy \and   
                INAF - Osservatorio Astronomico di Padova, Vicolo dell’Osservatorio 5, 35122, Padova, Italy \and
                Dipartimento di Fisica e Astronomia Galileo Galilei - Universit\`{a} di Padova, Vicolo dell'Osservatorio 2, 35122 Padova, Italy \and
                Fundación Galileo Galilei - INAF, Rambla José Ana Fernandez Pérez 7, E-38712 Breña Baja, TF - Spain \and
                INAF - Osservatorio Astronomico di Brera, via E. Bianchi 46, 23807 Merate (LC), Italy \and
                INAF - Osservatorio Astronomico di Capodimonte, Via Moiariello, 16, 80131 - NAPOLI}

   \date{Accepted and edited -- 20-12-2017}

 
  \abstract
  {We aim to investigate the presence of signatures of magnetic cycles and rotation on a sample of 71 early M-dwarfs from the HADES RV programme using high-resolution time-series spectroscopy of  the Ca II H \& K and H$\alpha$ chromospheric activity indicators, the radial velocity series, the parameters of the cross correlation function and the V-band photometry. We used mainly HARPS-N spectra, acquired over four years, and add HARPS spectra from the public ESO database and ASAS photometry light-curves as support data, extending the baseline of the observations of some stars up to 12 years.  We provide $\log_{10}(R'_\textrm{HK})$ measurements for all the stars in the sample, cycle length measurements for 13 stars, rotation periods for 33 stars and we are able to measure the semi-amplitude of the radial velocity signal induced by rotation in 16 stars.  We complement our work with previous results and confirm and refine the previously reported relationships between the mean level of chromospheric emission, measured by the $\log_{10}(R'_\textrm{HK})$, with the rotation period, and with the measured semi-amplitude of the activity induced radial velocity signal for early M-dwarfs. We searched for a possible relation between the measured rotation periods and the lengths of the magnetic cycle, finding a weak correlation between both quantities. Using previous $v$ sin $i$ measurements we estimated the inclinations of the star's poles to the line of sight for all the stars in the sample, and  estimate the range of masses of the planets GJ 3998 b and c (2.5 - 4.9 M$_{\oplus}$ and 6.3 - 12.5 M$_{\oplus}$), GJ 625 b (2.82 M$_{\oplus}$), GJ 3942 b (7.1 - 10.0 M$_{\oplus}$) and GJ 15A b (3.1 - 3.3 M$_{\oplus}$), assuming their orbits are coplanar with the stellar rotation.}

   \keywords{techniques:spectroscopy -- stars:late-type -- stars:low-mass -- stars:activity
               }

   \maketitle
%
\section{Introduction}

Since modern spectrographs reached the sub m s$^{-1}$ precision, stellar activity has become the biggest obstacle in the ongoing quest to find small mass planets via radial velocity (RV) measurements. Changes in the distribution of spots induce apparent Doppler shifts from less than one to dozens of metres per second, depending on the level of stellar activity \citep{Huelamo2008, Pont2011, Hatzes2013}. 

M-dwarfs are currently a case of special interest, as it is already possible to  detect potentially habitable Earth-like planets around them  \citep{Wright2016, AngladaEscude2016, Jehin2016, Gillon2017,AstudilloDefru2017, Masca2017,Masca2017c}, but stellar activity makes these detections rather difficult. The combination of surface inhomogeneities and stellar rotation easily mimic those of planetary origin \citep{Queloz2001, Bonfils2007,Boisse2011, Robertson2013,Howard2014, Masca2015,Newton2016,Vanderburg2016, Masca2017b}. For the case of M-dwarfs the amplitudes of these signals are comparable to those of rocky planets in the habitable zone of their stars \citep{Howard2014,Robertson2013,Masca2015,Newton2016,Masca2017b}. 

The determination of the rotation periods and magnetic cycles of the parent stars and their induced signals in the different activity indicators and radial velocity series is a crucial first step in discriminating true planetary signals from those induced by the stellar activity. In this paper we report the mean level of chromospheric activity  (measured as the $\log_{10}(R'_\textrm{HK}$), rotation periods, magnetic cycles and amplitudes of the rotation induced RV signals of the stars of the HArps-n red Dwarf Exoplanet Survey (HADES~\citep{Affer2016}). By using empirical relationships we also report the expected rotation periods and amplitudes of the rotation induced RV signals of those stars where a direct determination is not possible with the amount of available data.

\section{The HARPS-N M-dwarf radial velocity programme}
The HADES Radial Velocity Program is an extensive survey dedicated to the search of rocky extrasolar planets around M-dwarfs in the northern hemisphere. It is being conducted by an international consortium formed by two Spanish institutions -- the Instituto ~de ~Astrof\'{i}sica ~de ~Canarias (IAC) and the Institut ~d'Estudis ~Espacials ~de ~Catalunya (IEEC) -- and the Italian GAPS collaboration (Global ~Architecture ~of ~Planetary ~Systems, see~\citet{Covino2013} and ~\citet{Benatti2017}). The project has already led to the discovery of several planetary companions: a system of super-Earths around the star GJ 3998 \citep{Affer2016}, a super-Earth on the inner edge of the habitable zone of GJ 625 \citep{Masca2017c} and a super-Earth orbiting the active star GJ 3942 \citep{Perger2017b}. Observations have been conducted since the end of 2012 and the program is still ongoing. Up to date it has accumulated more than $3~500$ spectra of 71 targets \citep{Maldonado2017} obtained with HARPS-N at the Telescopio Nazionale Galileo. Some of the stars have previous observations performed with HARPS-S. In such cases we complemented our observations with all the spectra available in the HARPS ESO public data archive. We also used the ASAS survey to study the photometric time series of targets with declinations lower than +28º. Table~\ref{tab:star_parameters} shows the available data for all the stars under study.

\subsection{Spectroscopic data}

The two fibre-fed high-resolution \'{e}chelle spectrographs HARPS-N and HARPS \citep{Cosentino2012, Mayor2003} are installed at the 3.6 m Telescopio Nazionale Galileo (TNG) at the Roque de los Muchachos Observatory on La Palma (Spain) and at the 3.6 m ESO telescope in La Silla Observatory (Chile) respectively. Both instruments have a resolving power $R\sim 115\,000$ over a spectral range from $\sim$380 to $\sim$690 nm and have been designed to attain very high long-term radial velocity precision. Both are contained in vacuum vessels to limit spectral drifts due to temperature and air pressure variations, thus  ensuring their stability. HARPS and HARPS-N are equipped with their own pipeline providing extracted and wavelength-calibrated spectra, as well as RV measurements and other data products such as cross-correlation functions and their bisector profiles.

For our analysis, we used the extracted order-by-order wavelength-calibrated spectra produced by the HARPS and HARPS-N pipelines. In order to minimize the effects related to atmospheric changes and spectral throughput we created a spectral template for each star by co-adding every available spectrum and use the co-added spectrum to correct the order-by-order flux ratios for the individual spectra. We  corrected each spectrum for the Earth's barycentric radial velocity and the radial velocity of the star using the measurements given by the standard pipeline. We finally re-binned the spectra into a wavelength-constant step.  Using this  HARPS-N/S dataset, we expect to have high quality spectroscopic indicators  to monitor tiny stellar activity variations with  high accuracy. 

\subsection{Photometric data}
The All Sky Automated Survey (ASAS) \citep{Pojmanski1997} is an all sky survey in the $V$ and $I$ bands running since 1998 at Las Campanas Observatory, Chile. It has an angular resolution of $14"/pixel$ and an average accuracy of $\sim0.05$ mag per exposure. Best photometric results are achieved for stars with V $\sim$8-12, but this range can be extended  implementing  some quality control on the data. ASAS has produced light-curves for around $10^{7}$ stars at $DEC < 28^{\circ}$. The catalogue supplies ready-to-use light-curves with flags indicating the quality of the data. For this analysis we relied  only on  good quality data (grades `A' and `B' in the internal flags). Even after this quality control,  there are still some high dispersion measurements  which cannot  be explained by `regular' stellar behaviour. There are  some cases of extreme scatter in stars closer to the magnitude  limits of the survey and some stars that show behaviour compatible with stellar flares. As our data are not well suited for modelling fast events,  to remove  points affected by flares we iteratively rejected all measurements that deviate three standard deviations from the median value of the de-trended time series. The iteration was applied  until no more measurements were left  outside  these limits.

\section{Activity indicators and radial velocities}

To analyse stellar activity signals performed in this work we relied on two spectroscopic indicators -- the $S_{MW}$ index, based on the Ca II H\&K lines, and the H$\alpha$ index -- the full width half maximum (FWHM) of the cross correlation function (CCF), and the radial velocity (RV) data. For the calculation of the $S_{MW}$ index ~\citep{Noyes1984} we follow the procedure explained in \citet{Lovis2011} and \citet{Masca2015}. Then we used the $S_{MW}$  measurements to calculate the $\log_{10}R'_{HK}$ ~\citep{Noyes1984} by following the procedure used in ~\citet{Masca2015, Masca2016} to extend the index range of applicability to M-dwarfs. For the RV and FWHM measurements we followed the procedure explained in \citet{Masca2017, Masca2017c} to model the CCF. The HARPS and HARPS-N FWHM of the CCF show a trend with time \citep{Benatti2017b}. As our analysis relies only in periodic temporal variations we linearly detrended the FWHM time series. Due to the slow projected rotational velocity of most M-dwarfs in our sample, the bisector analysis becomes very complicated and it frequently does not offer any significant information \citep{Saar1997, Bonfils2007}. We performed the same rejection procedure stated in the previous section to remove  flare affected points and outliers. The cleaning procedure typically removed only a few points. 

\section{Time-series analysis of activity indicators}

We searched for periodic variability in the spectroscopic time series compatible with both stellar rotation and long term magnetic cycles. We computed the power spectrum using a generalized Lomb Scargle (GLS) periodogram \citep{Zechmeister2009} and if there is any significant periodicity we fitted the detected period, using a double-sinusoidal model at P and P/2 \citep{BerdyuginaJarvinen2005}, with the MPFIT routine \citep{Markwardt2009}. Then we repeated the same process in the residuals of the fit. Typically this allowed us to determine  the stellar rotation, and in some cases to unveil the presence of a sun-like cycle. In those cases where both quantities have been determined the final parameters come from a simultaneous fit of both signals. The significance of the periodogram peak was evaluated using the \citet{Cumming2004} modification of the \citet{HorneBaliunas1986} formula to obtain the spectral density thresholds for a desired false alarm level. Given the number of stars and datasets per star we opted the analytical expression instead of bootstrapping randomization for time-efficiency reasons. Our false alarm probability then gets defined as  $FAP = 1 - [1-P(z > z_{0})]^{M}$ where $P(z > z_{0})$ is the probability of $z$, the target spectral density, being greater than $z_{0}$, the measured spectral density, and $M$ the number of independent frequencies.  We explored a range of periods from one day up to a 50\% longer than the observation baseline of each individual target. The determination of incomplete cycles obviously involves large associated uncertainties. In those cases where the rotation or the cycle signal appear in more than one time series, the final result is the weighted mean between all the measurements and the error is the standard deviation of all the measurements. 

\begin{figure}[!h]
\includegraphics[width=9cm]{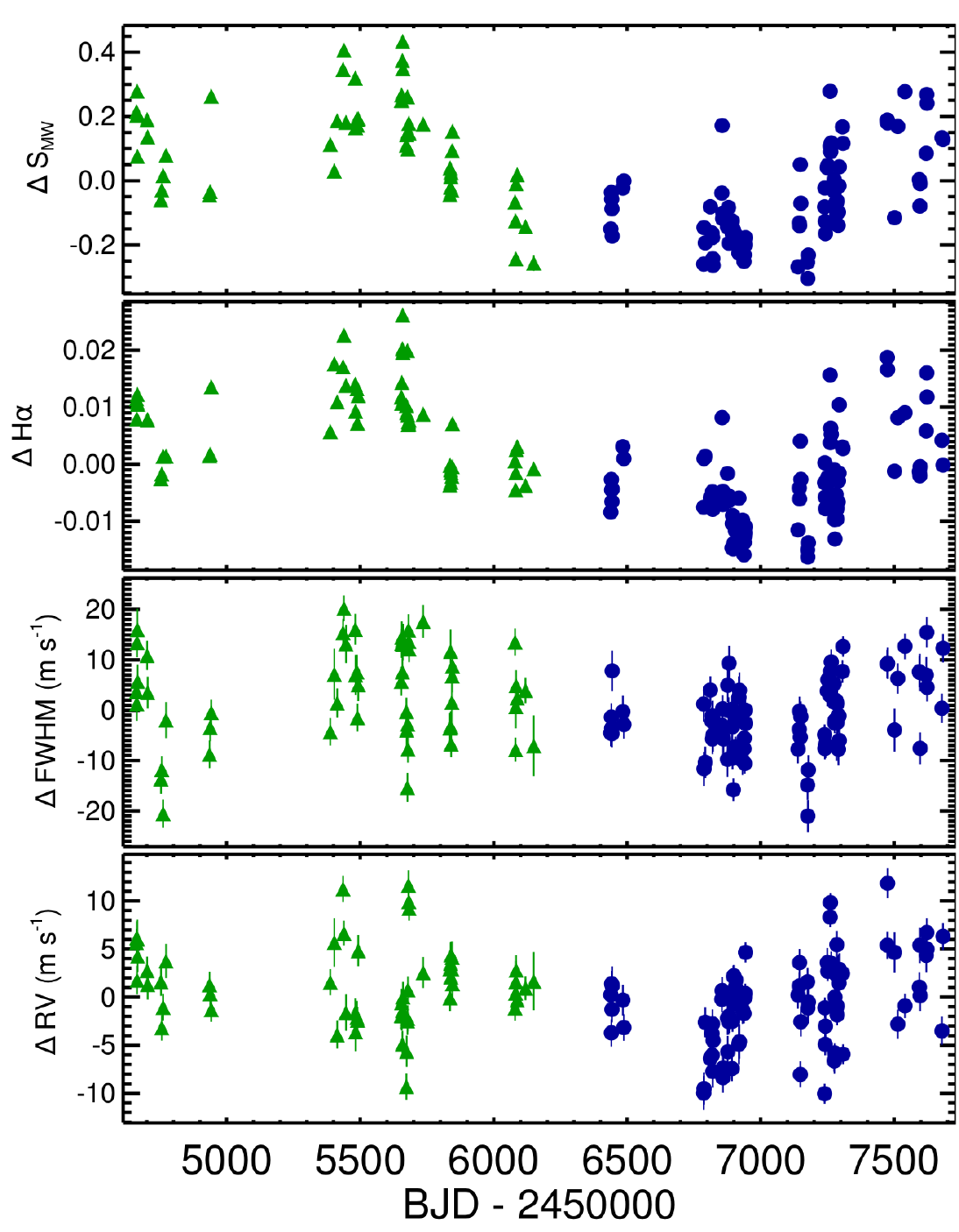}
\includegraphics[width=9cm]{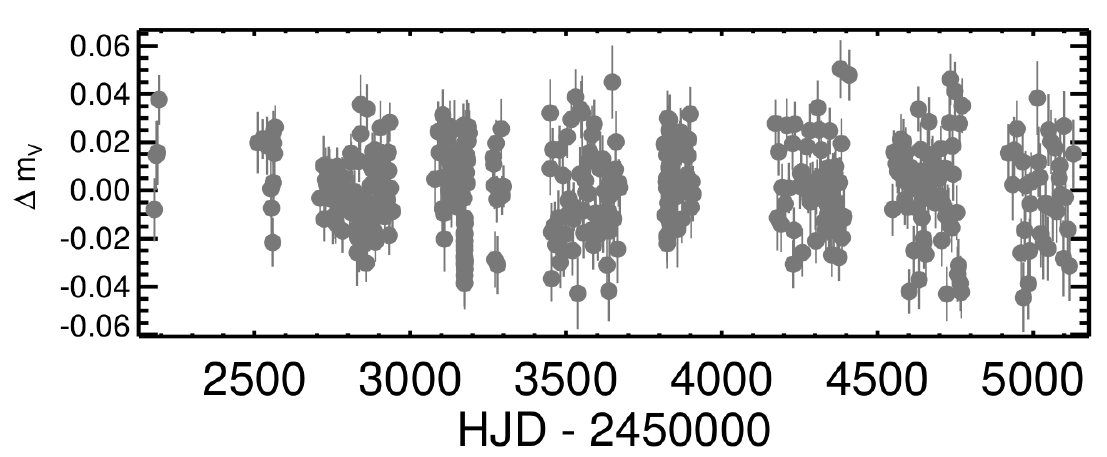}
\caption{Available data for the star GJ 740. From top to bottom: S$_{MW}$ index, H$\alpha$ index, FWHM, RV time series and ASAS light-curve. Blue dots show the HARPS-N data, green triangles show the HARPS data and grey dots show the ASAS data. }
\label{data_gj740}
\end{figure}

In order to illustrate the method,  we describe the case of the star GJ 740. GJ 740 is an M0.5 main sequence star located 10.9 pc away from the Sun \citep{vanLeeuwen2007}. It has a mass of 0.58 M$_{\odot}$, a radius of 0.56 R$_{\odot}$ and an effective temperature of 3845 K \citep{Maldonado2017}. It shows a mean $\log_{10}(R'_{HK}$) = --4.88 $\pm$ 0.04. Our dataset consists in 89 HARPS-N spectra obtained between January 2013 and September 2016, 53 HARPS spectra from the public ESO database obtained between 2008 and 2012 and 458 photometric exposures from the ASAS survey. Figure~\ref{data_gj740} shows the available GJ 740 data.

\begin{figure}[!h]
\includegraphics[width=9cm]{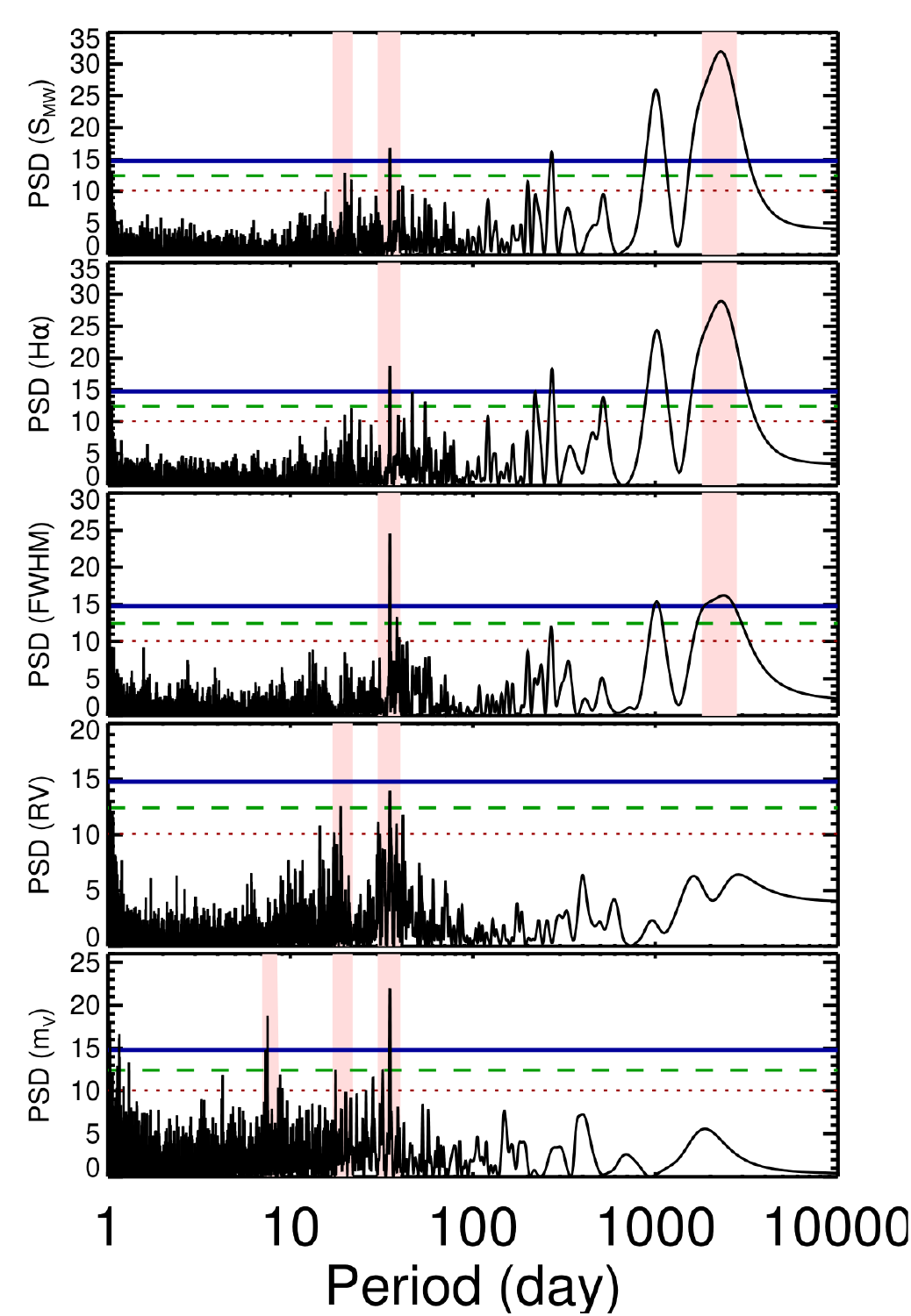}
\caption{GLS periodograms of the S$_{MW}$ index, H$\alpha$ index, FWHM, RV time series and ASAS m$_{V}$ light-curve of the star GJ 740. The shaded regions show the discussed signals. The horizontal lines show the 10\% (red dotted line), 1\% (green dashed line) and 0.1\% (blue solid line) FAP.}
\label{gls_gj740}
\end{figure}

\begin{figure}[!h]
\includegraphics[width=9cm]{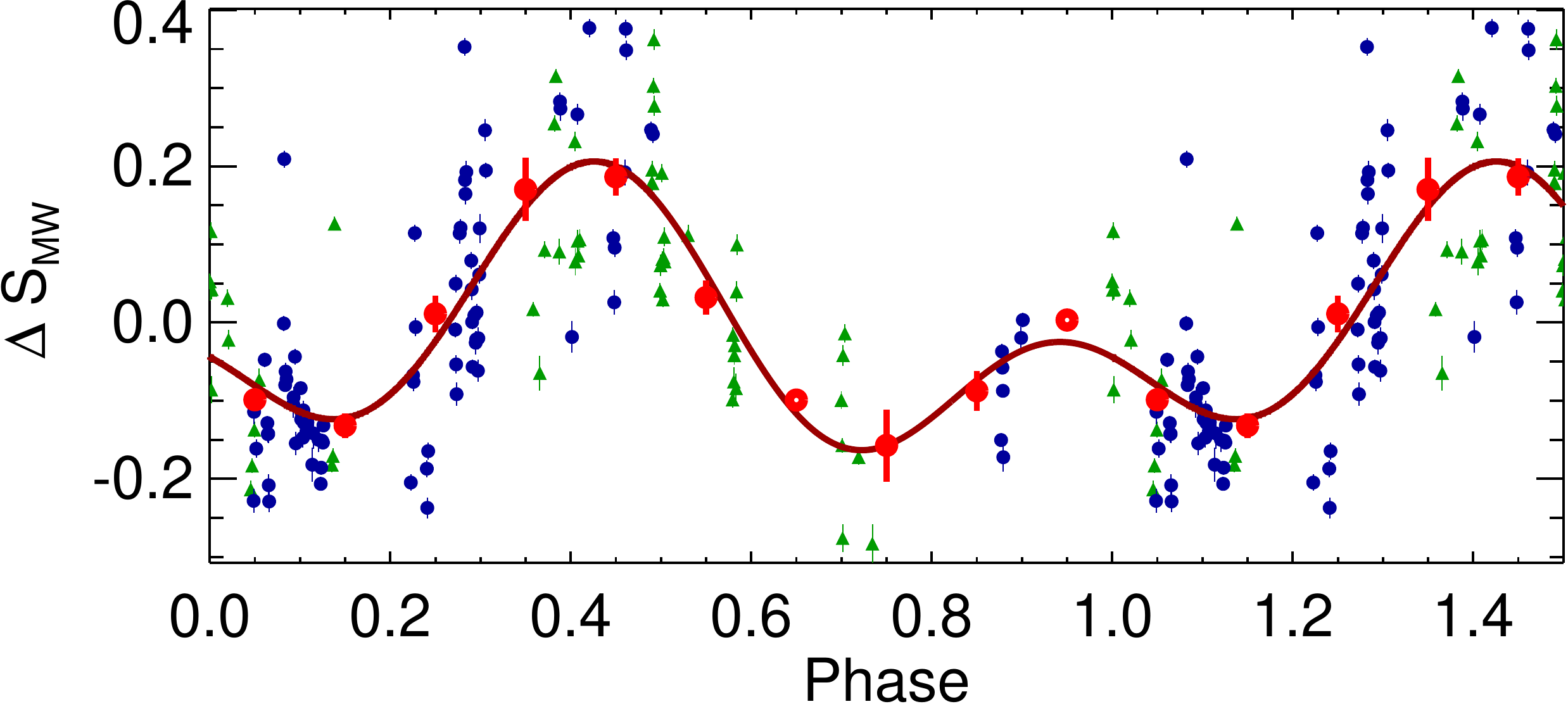}
\includegraphics[width=9cm]{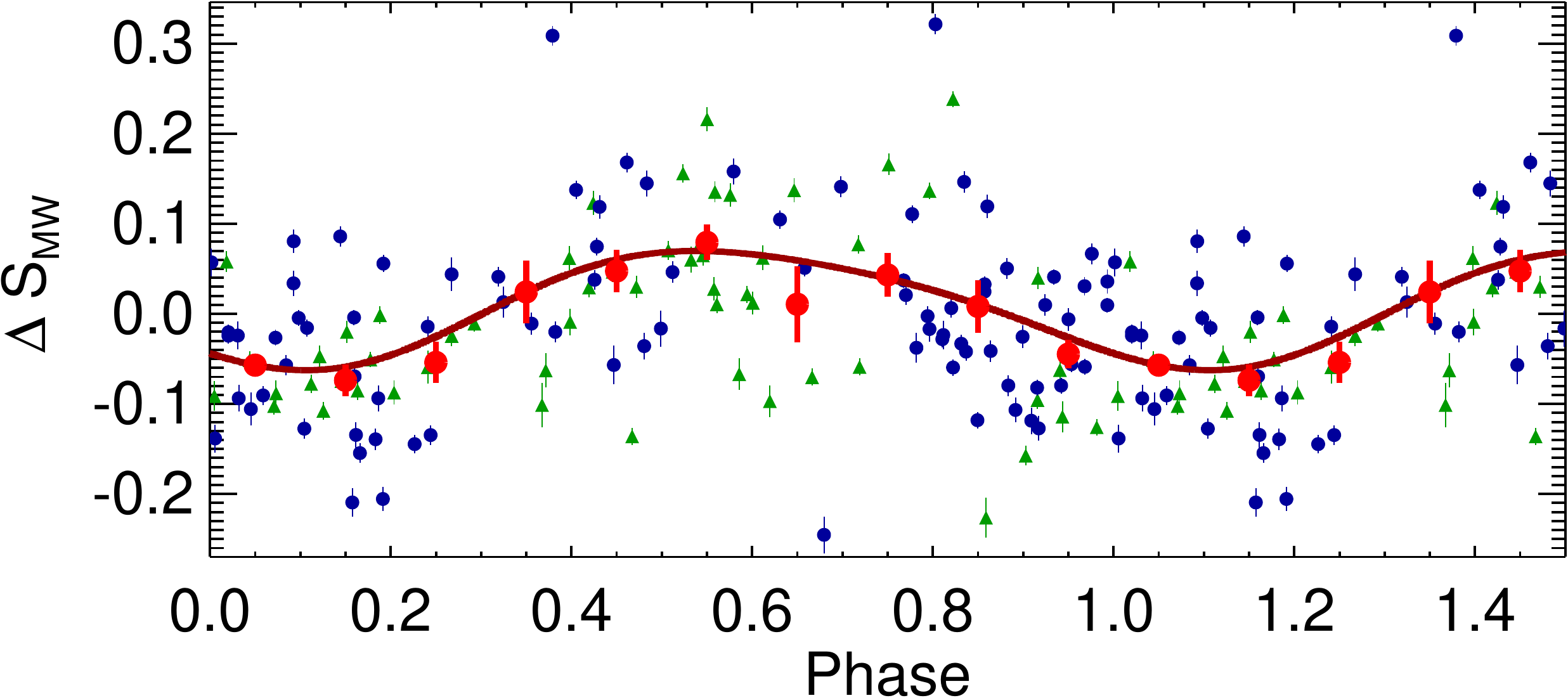}
\caption{Phase-folded fit of the  S$_{MW}$ index time series of the star GJ 740 using the 5.6 yr signal (top panel) and the 38.3 d signal (bottom panel)  using double-harmonic sinusoidal fits at P and P/2. Red dots show the mean measurement in each phase bin. In each case the contribution of the other signal has been subtracted. Blue dots show the HARPS-N data, green triangles show the HARPS data. }
\label{phase_smw}
\end{figure}

\begin{figure}[!h]
\includegraphics[width=9cm]{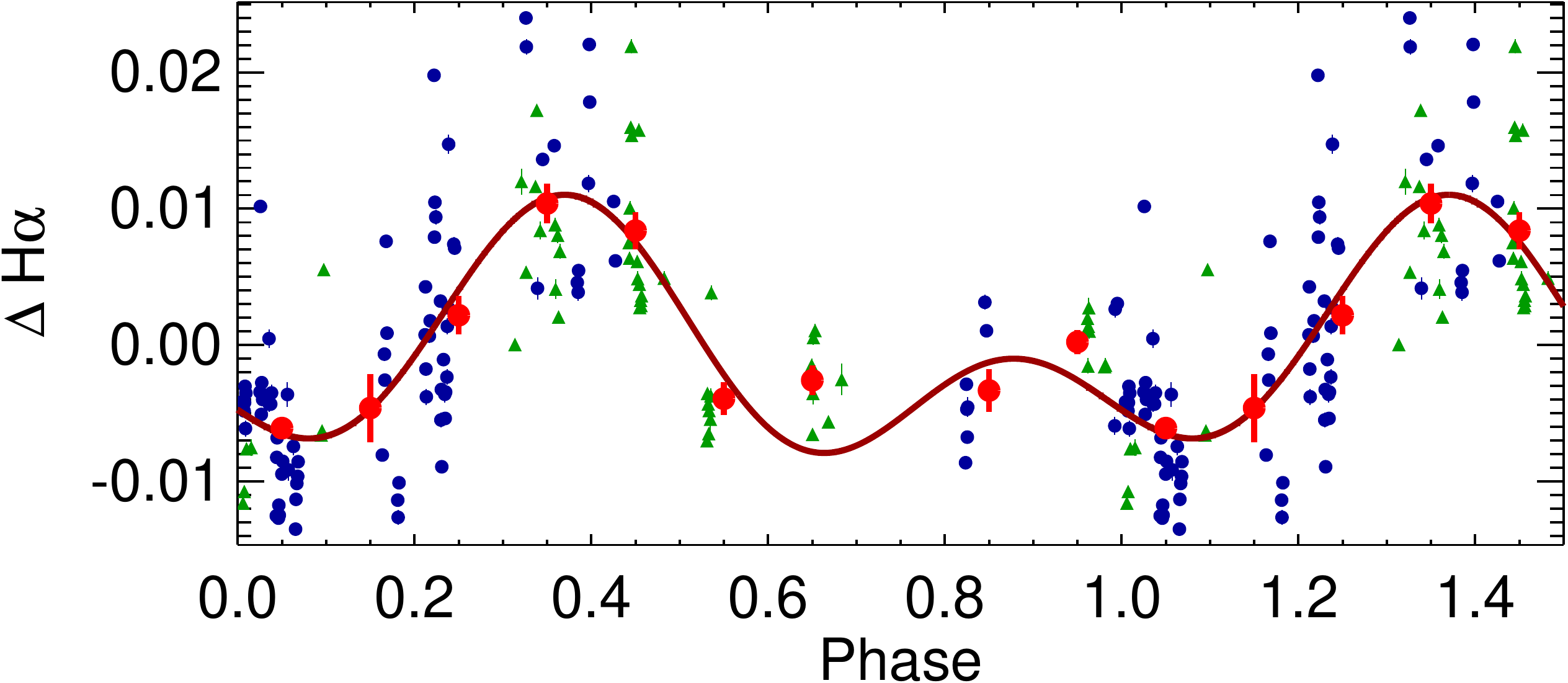}
\includegraphics[width=9cm]{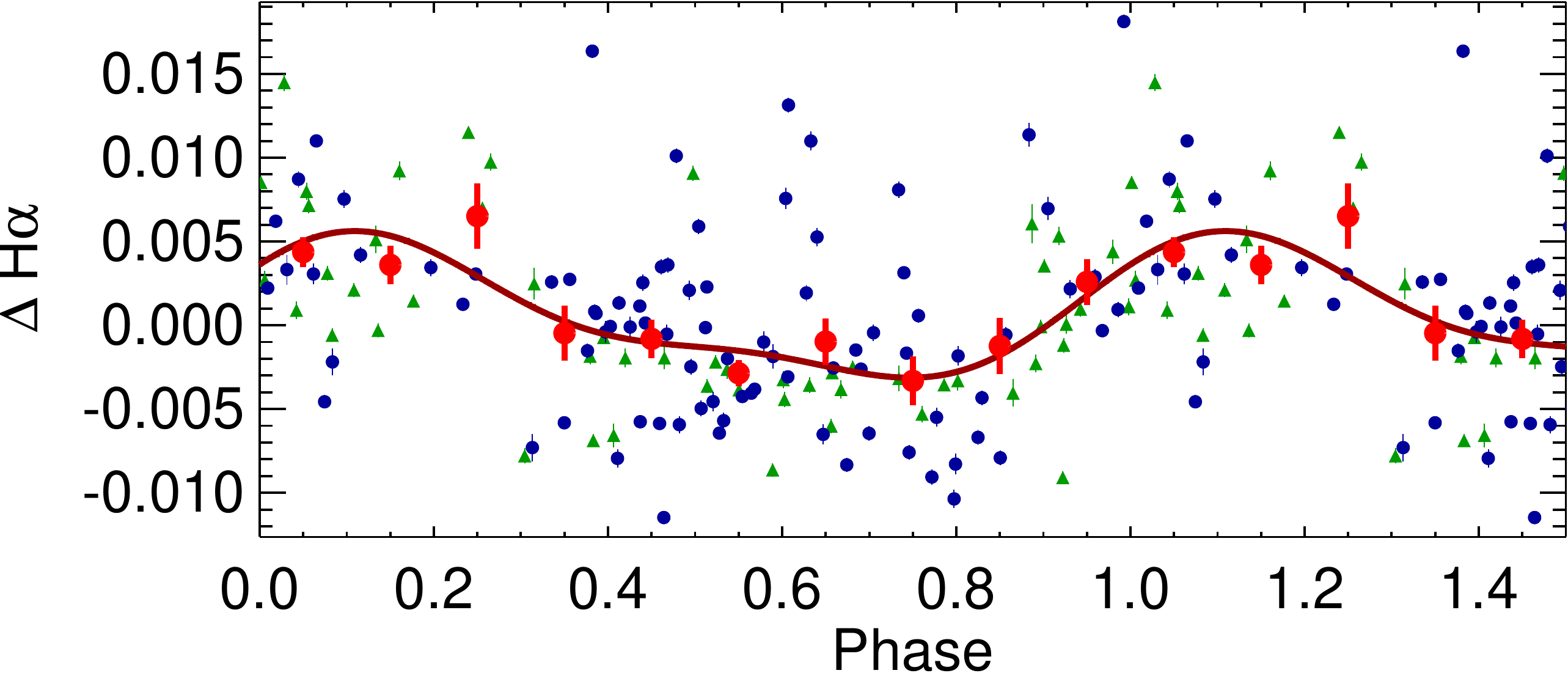}
\caption{Phase-folded fit of the H$\alpha$ index time series of the star GJ 740 using the 5.6 yr signal (top panel) and the 38.2 d signal (bottom panel) using doble-harmonic sinusoidal fits at P and P/2. Red dots show the mean measurement in each phase bin. In each case the contribution of the other signal has been subtracted. Blue dots show the HARPS-N data, green triangles show the HARPS data. }
\label{phase_ha}
\end{figure}

\begin{figure}[!h]
\includegraphics[width=9cm]{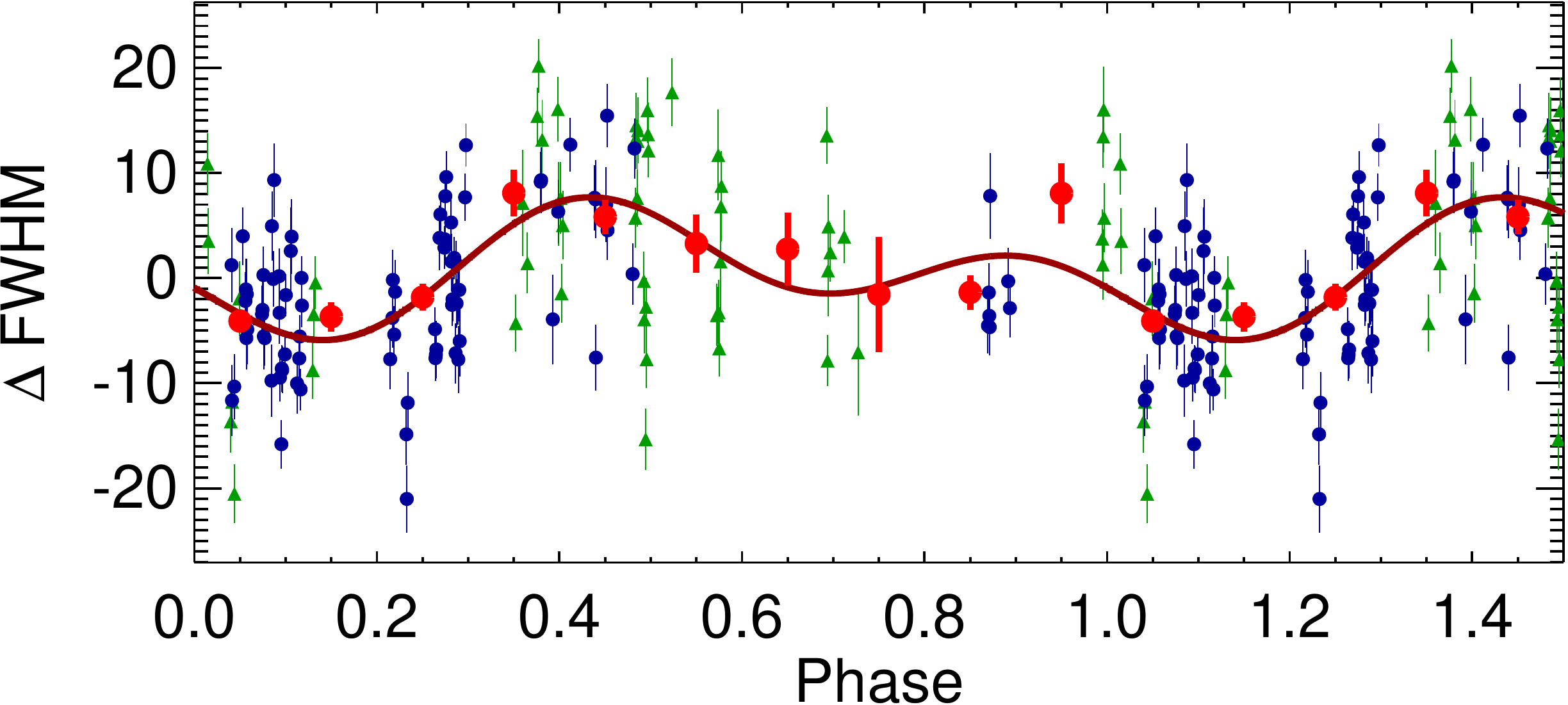}
\includegraphics[width=9cm]{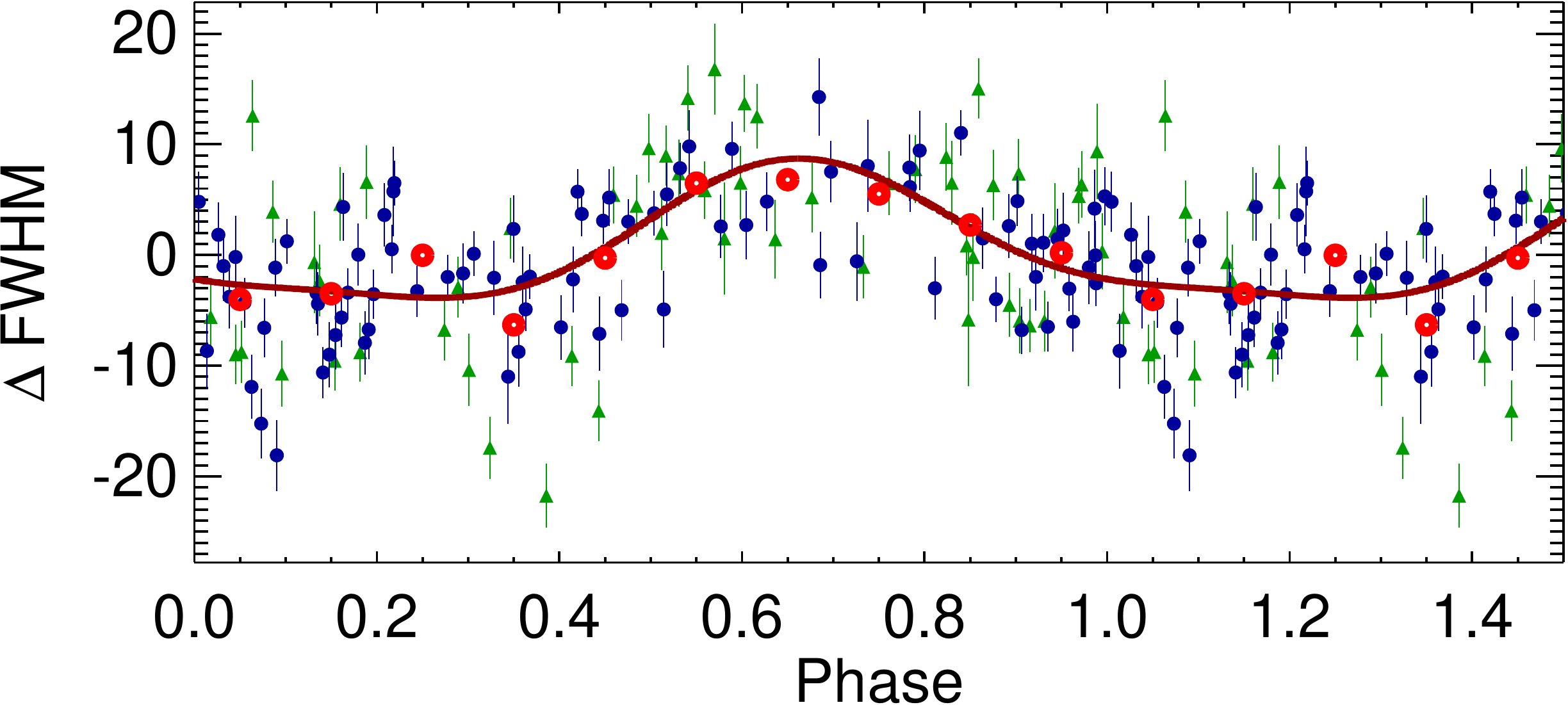}
\caption{Phase-folded fit of the FWHM index time series of the star GJ 740 using the 5.6 yr signal (top panel) and the 35.1 d signal (bottom panel)  using double-harmonic sinusoidal fits at P and P/2. Red dots show the mean measurement in each phase bin. In each case the contribution of the other signal has been subtracted. Blue dots show the HARPS-N data, green triangles show the HARPS data. }
\label{phase_fwhm}
\end{figure}

\begin{figure}[!h]
\includegraphics[width=9cm]{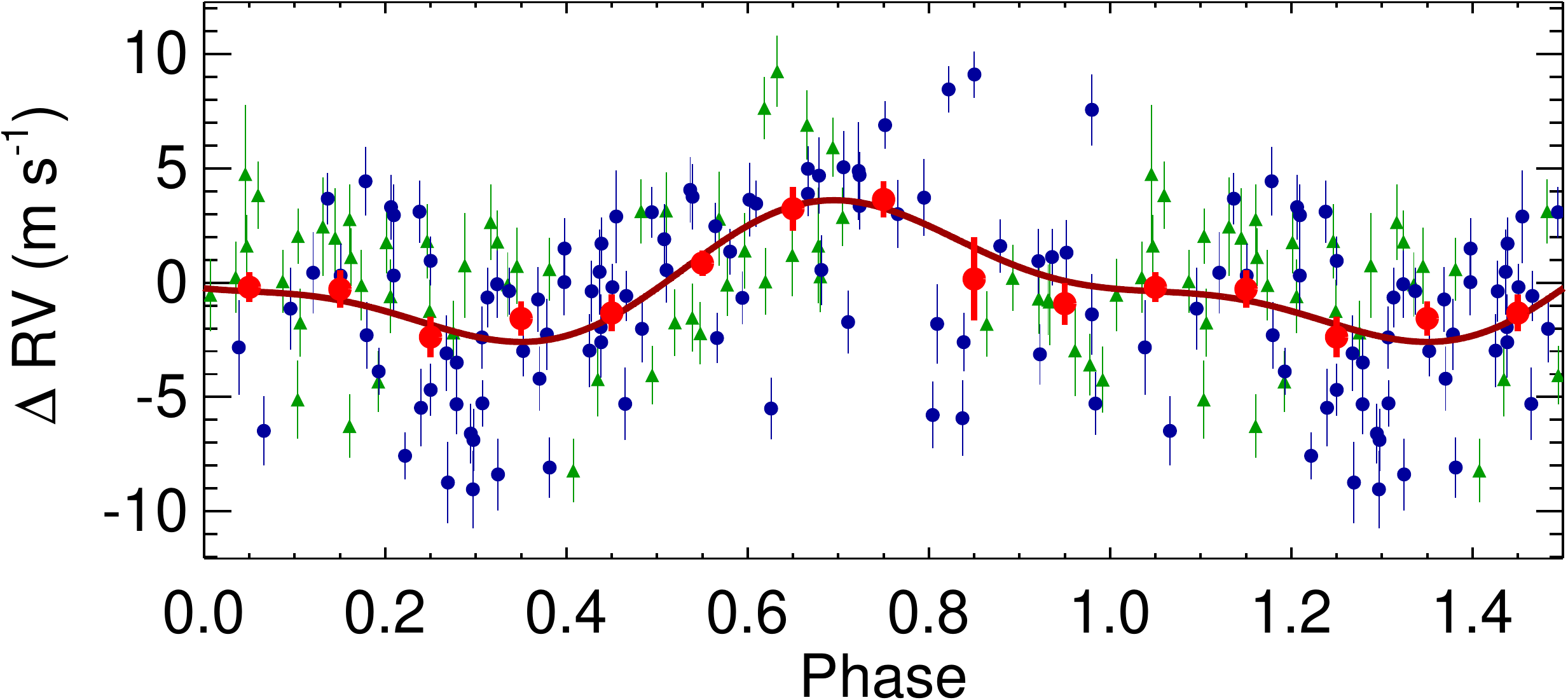}
\includegraphics[width=9cm]{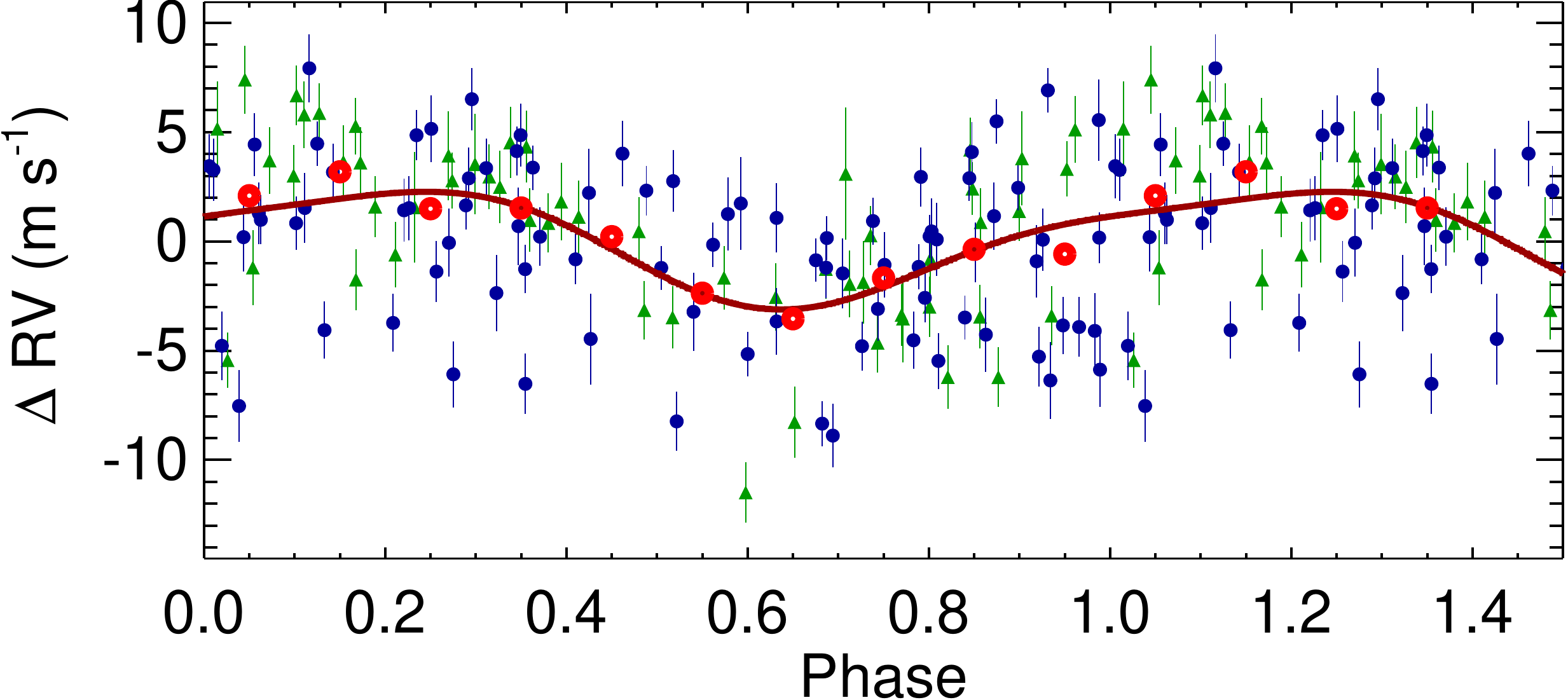}
\caption{Phase-folded fit of the  RV time series of the star GJ 740 using the 35.0 d signal (top panel) and the 17.6 d signal (bottom panel)  using double-harmonic sinusoidal fits at P and P/2. Red dots show the mean measurement in each phase bin. In each case the contribution of the other signal has been subtracted. Blue dots show the HARPS-N data, green triangles show the HARPS data. }
\label{phase_rv}
\end{figure}

\begin{figure}[!h]
\includegraphics[width=9cm]{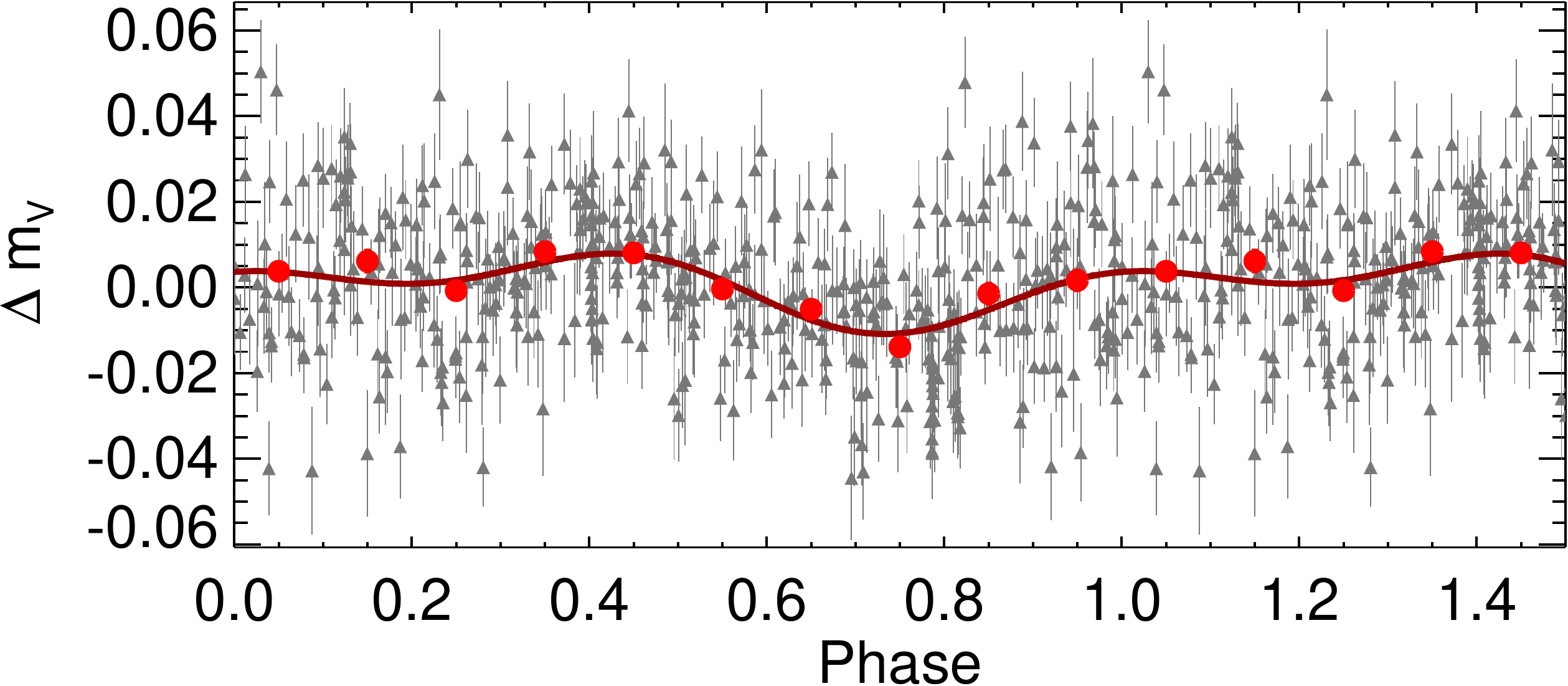}
\caption{Phase-folded fit of the ASAS m$_{V}$ time series of the star GJ 740 using the 35.2 d signal using a double-harmonic sinusoidal fit at P and P/2. Red dots show the mean measurement in each phase bin. }
\label{phase_vmag}
\end{figure}

An F-test \citep{Zechmeister2009b} shows that the probability of the variability of the different time series being explained by the error bars is negligible (much smaller than the 0.1\%). The GLS periodograms of the different time series reveal several periodic signals in common across many of the datasets (Fig.~\ref{gls_gj740}). The S$_{MW}$ index time series shows the presence of a 5.6 yr signal with a semi-amplitude of 0.165 with a FAP $\sim$ 0.1\%, a 38.3 d signal with a semi-amplitude of 0.061, and a barely significant excess of power at $\sim$ 19 d. The fit of the 38.3 d signal makes the $\sim$ 19 d power excess disappear, implying it is most likely the first harmonic of the main signal, caused by its non-sinusoidal shape. The phase-folded fits of both signals, the 5.6 yr and the 38.3 d, are shown in Fig.~\ref{phase_smw}. The H$\alpha$ index time series shows a very significant signal with a period of 5.6 yr and a semi-amplitude of 0.00658, and 38.2 d signal with a semi-amplitude of 0.00428. Figure~\ref{phase_ha} shows the phase-folded fits of both signals. The FWHM dataset shows also the same 5.6 yr signal, with a semi-amplitude of 2.58 m s$^{-1}$, and a 35.1 d signal with a semi-amplitude of 7.30 m s$^{-1}$. Figure~\ref{phase_fwhm} shows the phase-folded fits of both signals.  The RV time series shows two significant signals, at periods of 35.0 and 17.6 d with semi-amplitudes of 3.4 and 1.8 m s$^{-1}$ respectively with a clear non-sinusoidal shape (Fig.~\ref{phase_rv}). The ASAS light curve shows several seemingly periodic signals. The most prominent being a 35.2 d signal with a semi-amplitude of 11.1 mmag. Then a second significant signal at 7.5 d with a semi-amplitude of 3.5 mmag, and a power excess at $\sim$ 17 days. The final fit of the data (Fig.~\ref{phase_vmag}) shows that only the main signal at 35.2 d is a real signal, and the other two seem to be the first and second harmonics of the main signal, caused by the extreme non-sinusoidal shape of the signal. 

Using all the detected signals, and the information given by the different power excesses found in the periodograms, we conclude that GJ 740 shows a 5.6 years magnetic cycle and a rotation period of 36.3 days with a geometric distribution of the active regions that creates another signal at the first harmonic of the rotation period. The stellar rotation induces a RV signal of 3.4 $\pm$ 0.3 m s$^{-1}$ at the rotation period, and a secondary signal of 1.8 $\pm$ 0.3 m s$^{-1}$ at the first harmonic of the rotation period.   The cycle length, rotation period, and semi-amplitude of the star's activity-induced RV signal are very consistent with previous measurements in early M-dwarfs \citep{Masca2015, Masca2016, Masca2017b}.

Table~\ref{tab:results} shows the results obtained for those 33 stars. The final  period measurements are the weighted mean of the measurements obtained from the different indicators, the uncertainty adopted is the standard deviation of all the measurements. The global false alarm probability is the combined probability of the null hypothesis being true given by a Fisher's combined probability test  \citep{Fisher1925}, using the individual false alarm probabilities of the different detections. The rotation period measurements are in good agreement with the pooled variance analysis of \citet{Scandariato2017} for the stars in common.

\section{Discussion}

\begin{figure}[!h]
\includegraphics[width=9cm]{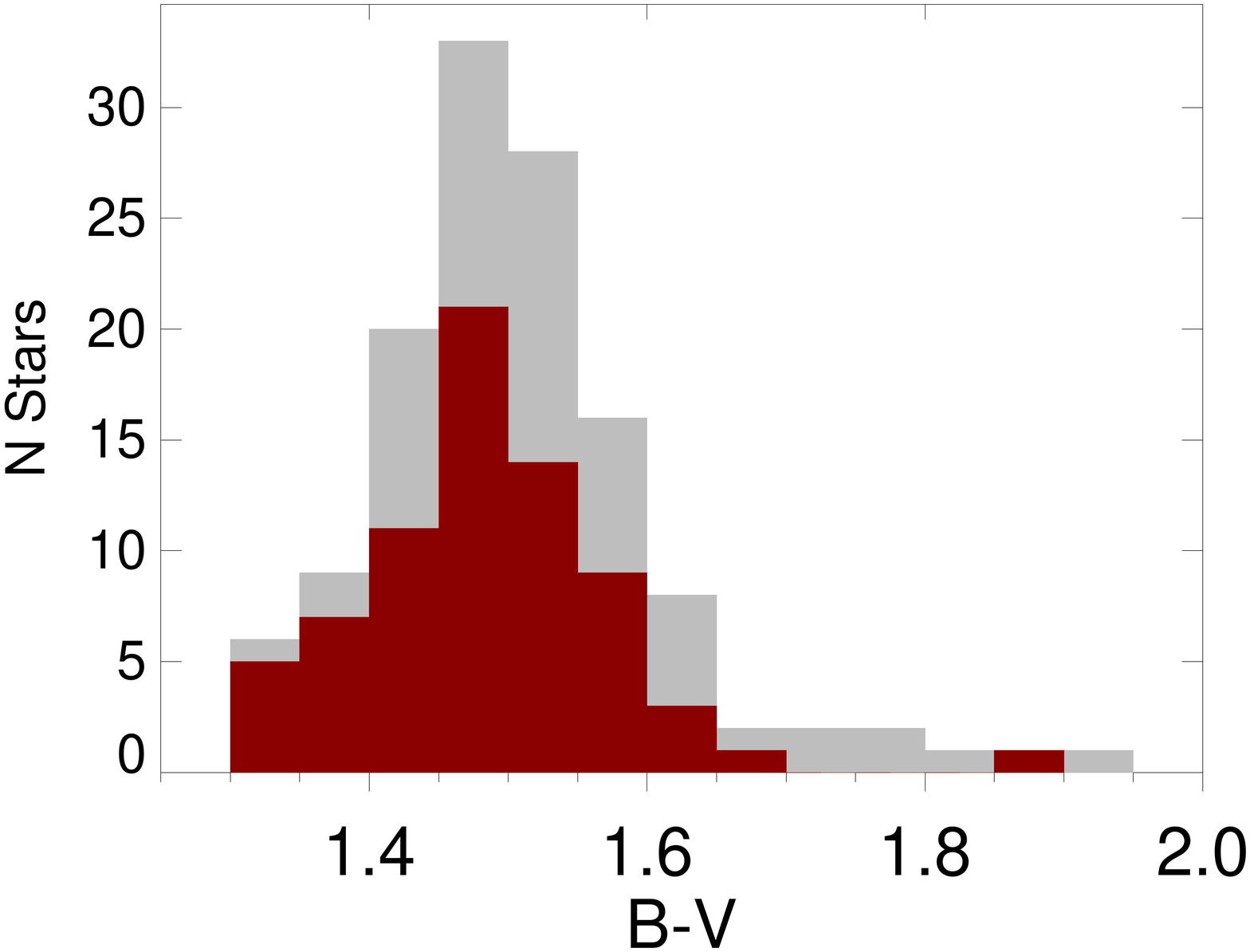}
\caption{Distribution of the color B-V of our sample. The grey area shows the full sample used in this discussion while red area shows the color of the stars newly analysed in this work. }
\label{bv_hist}
\end{figure}

The previous analysis of variability in spectroscopic and photometric time series and radial velocities  provided a collection of new measurements of chromospheric activity levels for 71 stars, rotation periods of 33 stars, magnetic cycle lengths for 13 stars -- along with lower limits of the cycle length for other five stars -- and RV signals induced by rotation for 16 stars. In order to put these results in a broader context and perform a better statistical analysis we merged our dataset with M-dwarfs included in the datasets of \citet{Robertson2013, Masca2015, Masca2016, Masca2017b} and \citet{Astudillo2017}. The combined dataset consists of 129 measurements of chromospheric activity levels, 90 rotation period measurements, 47 magnetic cycles, and 35 measurements of RV signals induced by rotation in M-dwarfs of early to mid spectral types. Table~\ref{tab:bv_stats} and Fig.~\ref{bv_hist} show the distribution of the color B-V along our sample. It peaks at B-V $\sim$ 1.5 with most stars distributed between B-V $\sim$ 1.4 and B-V $\sim$ 1.6.

\begin {table}
\begin{center}
\caption {Typical B-V color of the stars under study. \label{tab:bv_stats}}
    \begin{tabular}{  l l  l  l l } \hline
 Dataset   & N &  Mean & Median & $\sigma$  \\ 
        \hline

M0 - M6 & 129   & 1.50 & 1.50 & 0.10 \\ 
M0 - M3         & 114 & 1.49 & 1.49 & 0.09 \\
M3.5 - M6       & 15 & 1.60 &  1.61 & 0.12 \\\\
        
M0 - M3 (HADES) & 72 & 1.48 & 1.48 & 0.09 \\

 \hline
\end{tabular}  
\end{center}
\end {table}

\subsection{Distribution of $\log_{10}R'_{HK}$}\label{sect_rhk}

\begin{figure}[!h]
\includegraphics[width=9cm]{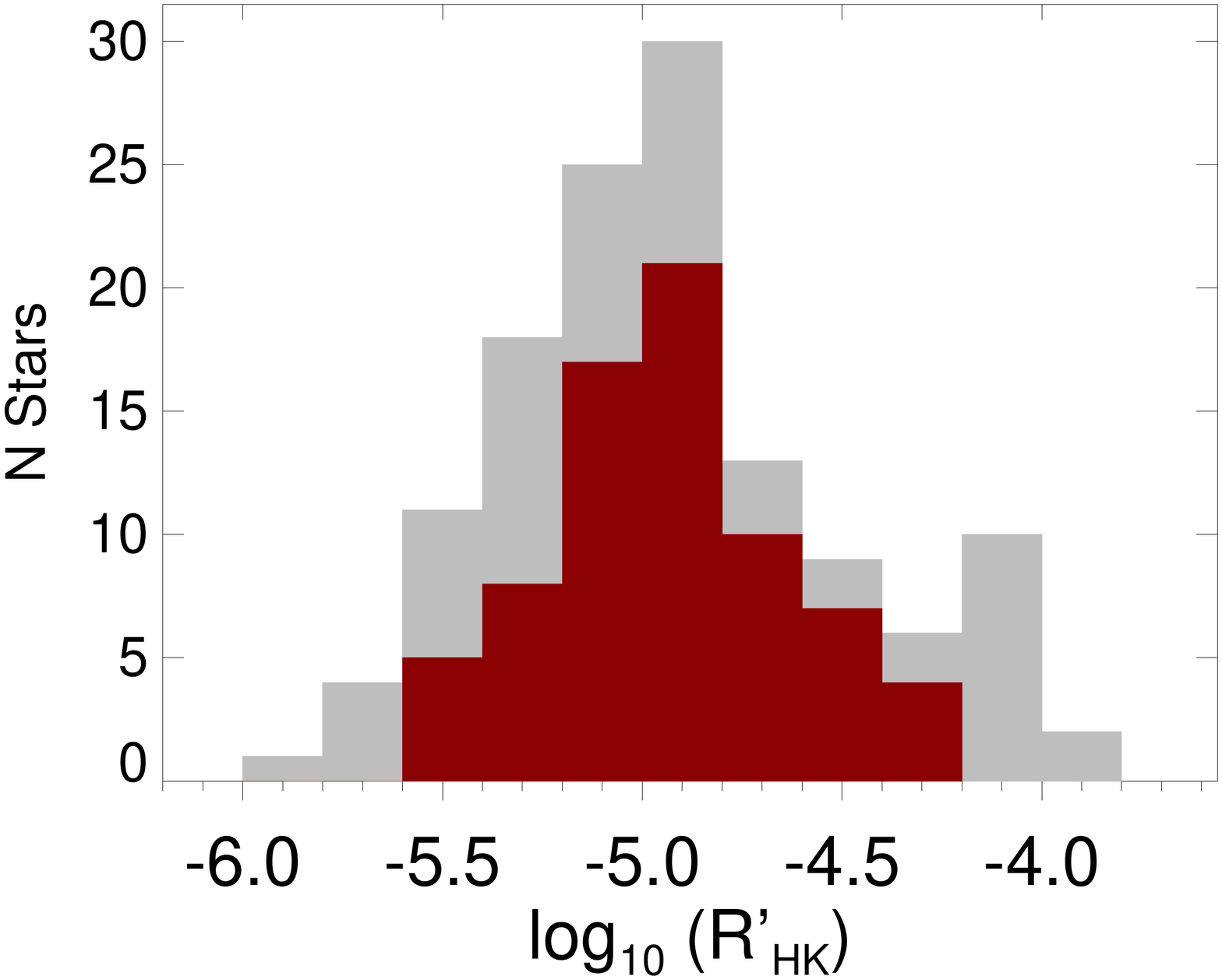}
\includegraphics[width=9cm]{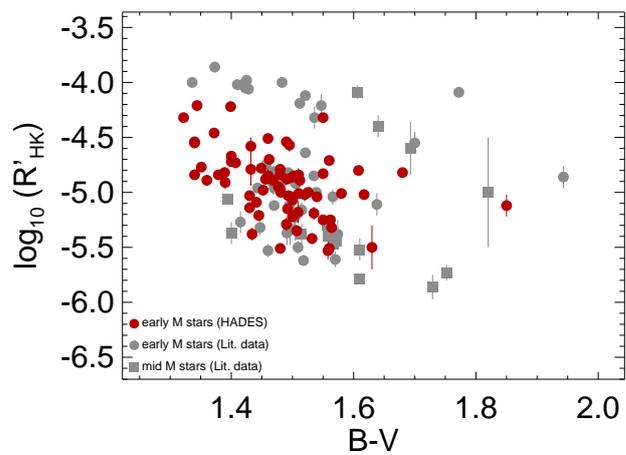}
\caption{\textbf{Top panel:} Distribution of the level of chromospheric activity, measured as the $\log_{10}R'_{HK}$, for the stars in our sample. The grey area shows the full sample, while the red highlights the fraction of new measurements. \textbf{Bottom panel:} Level of chromospheric activity, measured as the $\log_{10}R'_{HK}$, against the color B-V of the stars. Early M stars refer to M0 to M3 stars, mid M stars are M 3.5 to M6 stars. }
\label{bv_rhk}
\end{figure}

The available number of measurements of  $\log_{10}R'_{HK}$ for M-dwarfs in the literature is rather small. The sample used in this discussion, which includes the newly analysed stars and the measurements from \citet{Masca2015, Masca2016, Masca2017b} and \citet{Astudillo2017}, is one of the largest compilation of $\log_{10}R'_{HK}$ measurements for M-dwarfs in the literature. This sample shows rather low levels of chromospheric activity, as expected for a planet-hunting sample, with the mean level being very close to the Sun's mean activity level ($\log_{10}R'_{HK}$ $\sim$ -- 4.9) and the least active stars reaching almost a $\log_{10}R'_{HK}$ $\sim$ -- 6.0. These low levels of chromospheric emission are very uncommon in earlier type stars, with the most quiet FGK-type stars sitting around $\log_{10}R'_{HK}$ $\sim$ -- 5.2 (see \citet{Noyes1984, Baliunas1995, Lovis2011, Masca2016}). The upper limit of the distribution on the other hand is very similar to the upper envelope of FGK-type stars, reaching  levels slightly higher than -- 4.0. Our sample shows a correlation coefficient of --0.36 between the color B-V and the mean activity level $\log_{10}R'_{HK}$. Table~\ref{tab:rhk_stats} and Fig.~\ref{bv_rhk} show the distribution of the mean levels of chromospheric activity, measured as the $\log_{10}R'_{HK}$, for the stars in our sample. 

\begin {table}
\begin{center}
\caption {Typical level of chromospheric activity of the stars under study, measured as the $\log_{10}R'_{HK}$  \label{tab:rhk_stats}}
    \begin{tabular}{  l l  l  l l } \hline
Dataset    & N &  Mean & Median  & $\sigma$ \\ 
        \hline

M0 - M6 & 129   & -- 4.91 & --4.92 & 0.44 \\ 
M0 - M3         & 114 & -- 4.87 &  -- 4.89 & 0.41 \\
M3.5 - M6               & 15 & -- 5.23 &  -- 5.38 & 0.51 \\
\\
        
M0 - M3 (HADES) & 72 & -- 4.92 & -- 4.89 & 0.30 \\

 \hline
\end{tabular}  
\end{center}
\end {table}

\subsection{Rotation period distribution} \label{sect_rot}

\begin{figure}[!h]
\includegraphics[width=9cm]{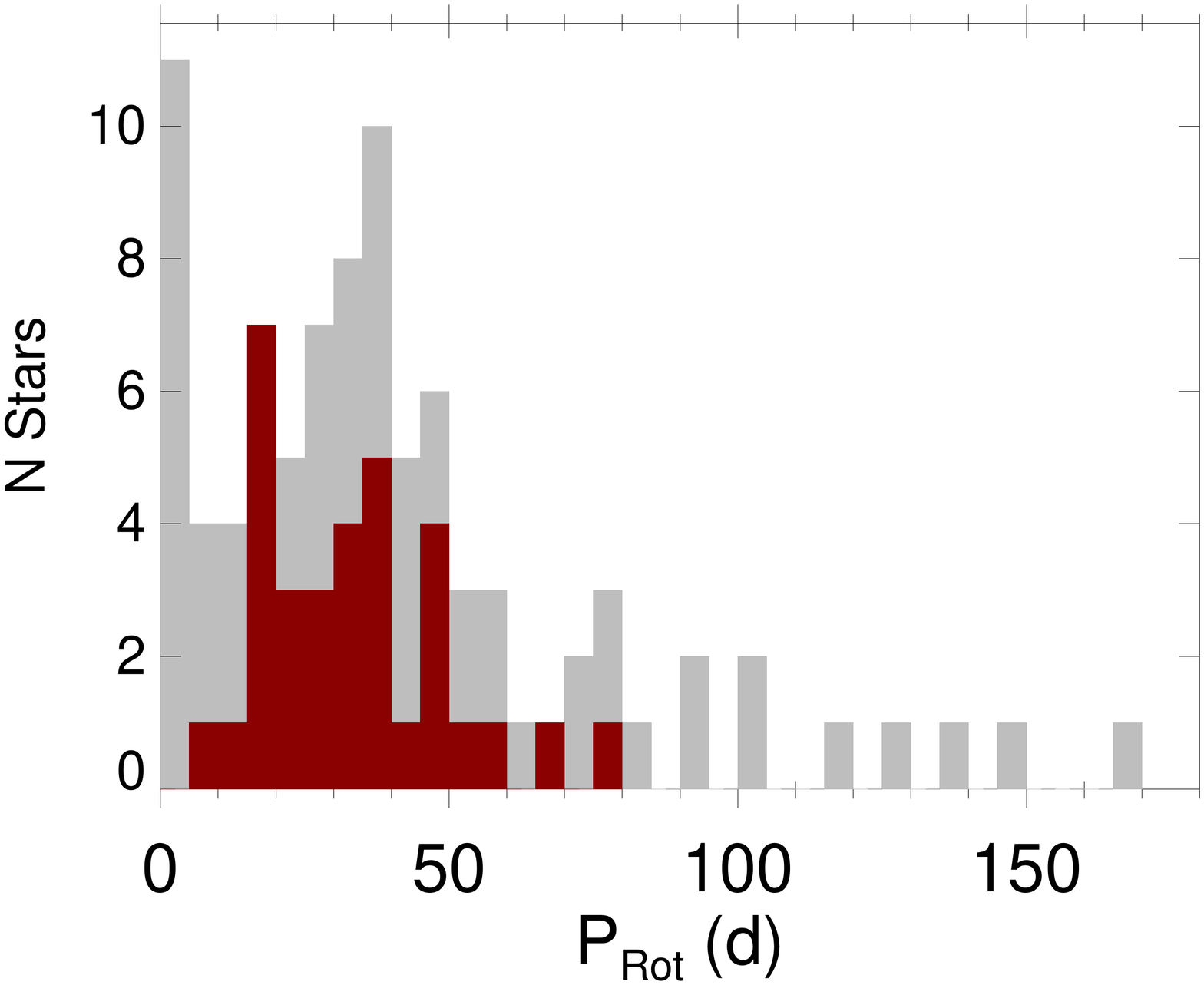}
\includegraphics[width=9cm]{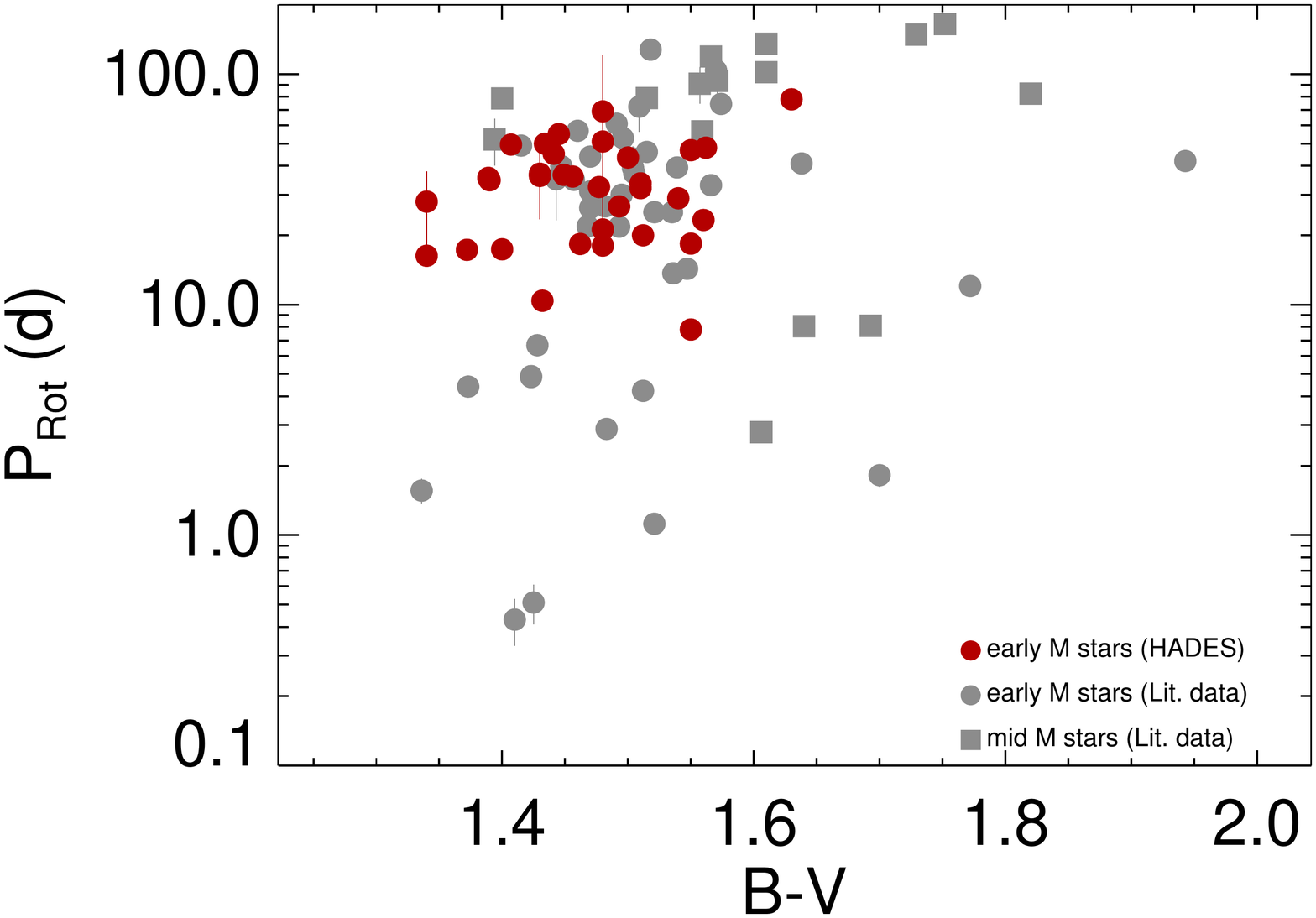}
\caption{\textbf{Top panel:} Distribution of the rotation periods for the stars in our sample. The grey area shows the full sample, while the red hightlight the fraction of new measurements. \textbf{Bottom panel:} Rotation period against the color B-V of the stars. Early M stars refer to M0 to M3 stars, mid M stars are M 3.5 to M6 stars.}
\label{bv_rot}
\end{figure}

Figure~\ref{bv_rot} shows the distribution of rotation periods for the stars in our sample and Table~\ref{tab:rot_stats} the typical periods and the measured scatter. Typical main sequence M-dwarfs are known to be rather slow rotators \citep{McQuillan2013,Newton2016}, and our data supports this. We measured a typical rotation period of $\sim$ 33 days for early M-dwarfs and of $\sim$ 81 days for mid M-dwarfs. Typical rotation periods for G- and K-type field dwarfs would be around 20 and 27 days respectively \citep{Masca2016}. The upper envelope of the distribution of periods shows the rotation periods increasing with redder stars (lower masses), going from $\sim$ 30 days for very early M-dwarfs to periods longer than 100 days for mid M-dwarfs, with the slowest rotators being also the stars that show the lowest mean $\log_{10}R'_{HK}$ $\sim$ levels. Our sample shows a correlation coefficient of 0.25 between the color B-V and the rotation period measurements. Figure~\ref{bv_rot} shows also a split between two groups, of fast and slow rotators, which probably points to a difference in age of the different groups of stars. Our sample is biased towards low activity stars (as most exoplanet surveys are), but our rotation period measurements and their distributions are consistent with the quiet branch of M-dwarfs discussed in previous works \citep{McQuillan2013, Newton2016, Masca2016, Astudillo2017}.

\begin {table}
\begin{center}
\caption {Typical rotation periods for the M-dwarfs under study.\label{tab:rot_stats}}
    \begin{tabular}{  l l  l  l l } \hline
Dataset    & N &  Mean & Median  & $\sigma$\\ 
        &       &  (d) & (d) &(d) \\\hline
M0 - M6 & 90    & 40 & 35 &  34\\ 
M0 - M3         & 75 & 33 &  32 & 23 \\
M3.5 - M6       & 15 & 81 &  82 & 50 \\\\
        
M0 - M3 (HADES) & 33 & 33 & 32 & 16 \\

 \hline
\end{tabular}  
\end{center}

\end {table}

\subsection{Cycle distribution}

\begin{figure}[!h]
\includegraphics[width=9cm]{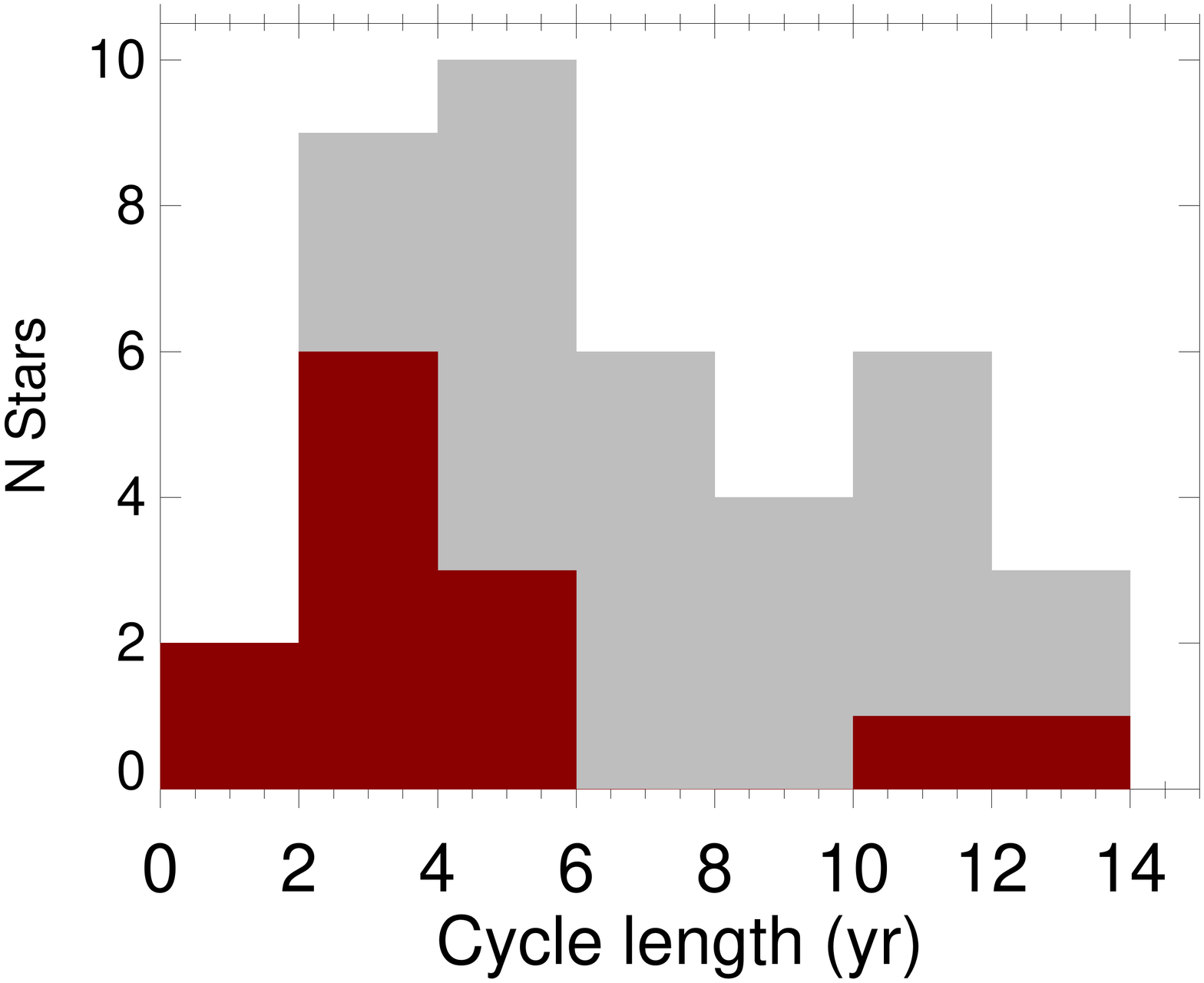}
\includegraphics[width=9cm]{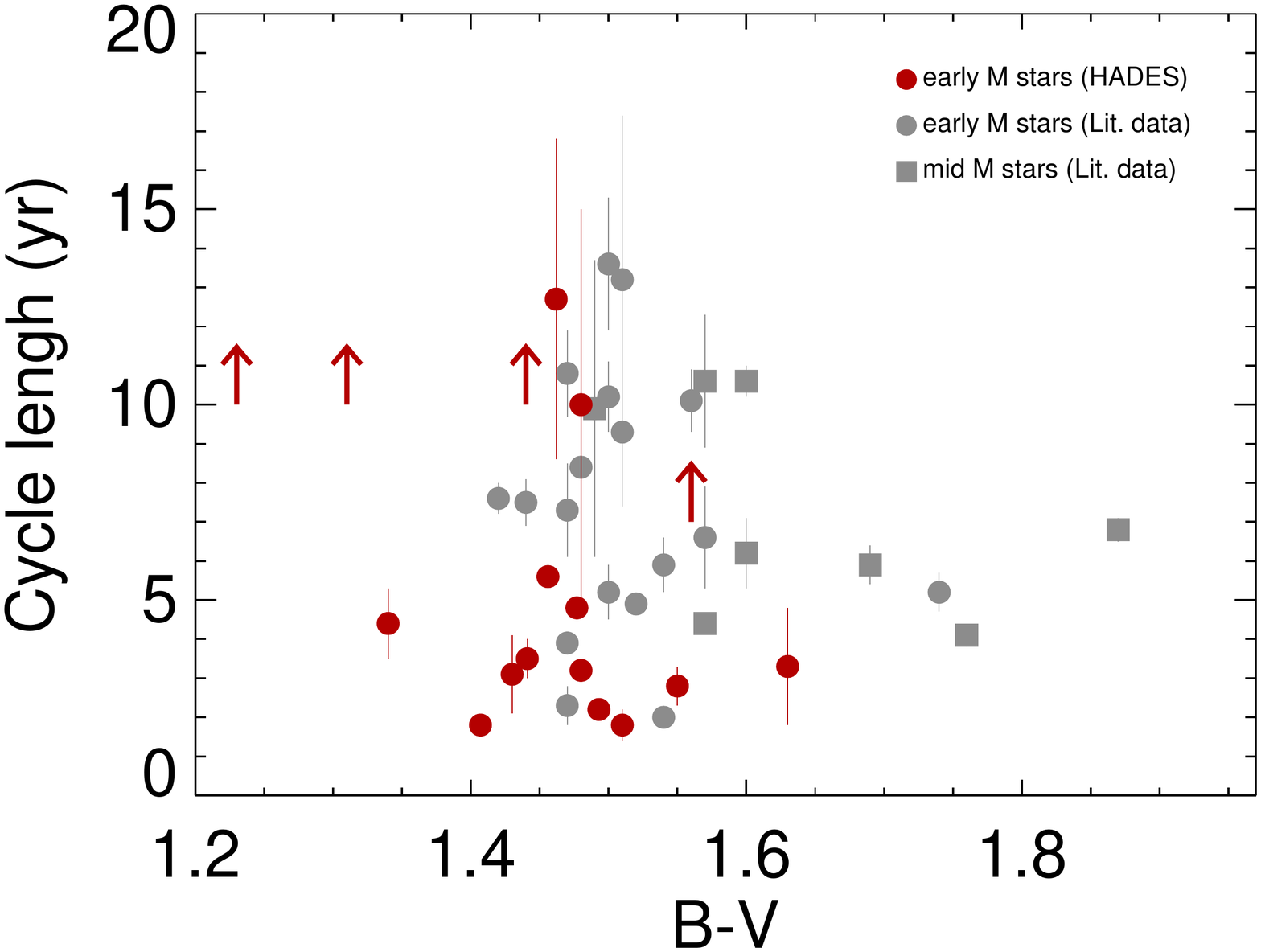}
\caption{\textbf{Top panel:} Distribution of the cycle lengths for the stars in our sample. The grey area shows the full sample, while the red hightlight the fraction of new measurements. \textbf{Bottom panel:} Cycle lengths against the color B-V of the stars. Vertical arrows indicate the lower limit of the periods for those cases where the presence of a cycle is detected but the period could not be determined.}
\label{bv_cyc}
\end{figure}

Figure~\ref{bv_cyc} shows the distribution of cycles by length and spectral type and Table~\ref{tab:cyc_stats} shows the main  statistics of the typical cycles for the stars in our sample. Cycle lengths peak at two to six years and then the number of detections decreases steadily with the longest cycles detected at almost 14 years. For the cycle measurements we are clearly limited by the time span of the observations. Long cycles (longer than approximately three to four years) are not measurable by our campaign on its own, they always require support data (HARPS or ASAS data), and even in these cases there are not many stars with a long enough baseline to draw significant conclusions over the presence of long cycles.  
 
\begin {table}
\begin{center}
\caption {Typical length of the measured magnetic cycles in the M-dwarfs under study.\label{tab:cyc_stats}}
    \begin{tabular}{  l l  l  l l } \hline
Dataset   & N &  Mean  & Median & $\sigma$  \\ 
        &       &  (yr) & (yr) &(yr) \\\hline
M0 - M6 & 45    & 6.5 & 5.9 &  3.3 \\ 
M0 - M3         & 36 & 6.3 &  5.2 & 3.4 \\
M3.5 - M6       & 9 & 7.5 &  6.8 & 2.6 \\
\\
        
M0 - M3 (HADES) & 13 & 4.7 & 3.3 & 3.3 \\

 \hline

\end{tabular}  
\end{center}

\end {table}

\subsection{Activity-rotation relation}

\begin{figure*}[!h]
\includegraphics[width=18cm]{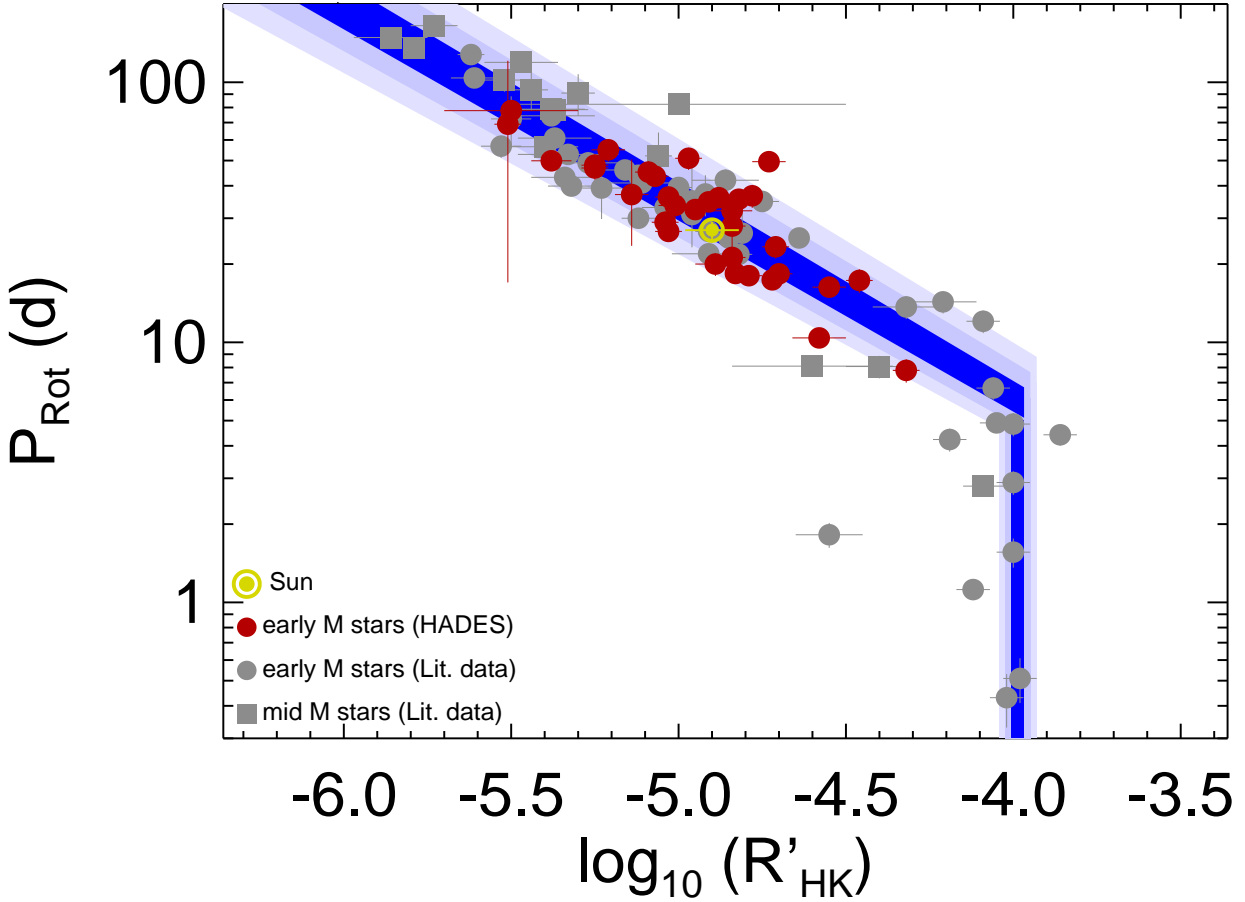}
\caption{Rotation period vs. chromospheric activity level $\log_{10}R'_{HK}$. The shaded region, going from darker to lighter blue, shows the 1,2,3 $\sigma$ confidence intervals of Eq.~\ref{eq_rhk_period} using the parameters of Table~\ref{tab:rhk_period_tab}, with the saturation level at $\log_{10}R'_{HK}$ -- 4.0 $\pm$ 0.1. Early M stars refer to M0 to M3 stars, mid M stars are M 3.5 to M6 stars. The Sun's value is included as a reference point.} 
\label{rhk_rot}
\end{figure*}

Activity-rotation relationships have been known for a long time ~\citep{Noyes1984, Duncan1991, Mamajek2008}. Many different indexes have been used depending on the stellar type and age. For main sequence FGK-type stars the relation between the $\log_{10}R'_{HK}$ and the rotation period (usually through the Rossby number) has been very widely used since it was introduced by \citet{Noyes1984}. \citet{Masca2015} showed that M-dwarfs could also fit a very similar relationship using a small sample of early and mid-type M-dwarfs. That claim was later supported by \citet{Masca2016} and \citet{Astudillo2017}, the latter including also a group of fast rotators that showed a saturation limit at $\log_{10}R'_{HK}$ $\sim$ -- 4. Figure~\ref{rhk_rot} shows the measurements from \citet{Masca2015, Masca2016, Masca2017b} and \citet{Astudillo2017} along with the 33 new measurements from this work. The 90 measurements together confirm and strengthen the proposed relationship, for stars with $\log_{10}R'_{HK} \leq -4.1$. The Spearman's correlation coefficient between the $\log_{10}R'_{HK}$ and the rotation period is -- 0.93, with a probability of being spurious smaller than 10$^{-38}$. The relationship can be written as:

\begin{equation}
\begin{split}
  \log_{10}(P_{rot})= A + B \cdot \log_{10}R'_{HK} 
\end{split}
\label{eq_rhk_period}
,\end{equation}

\noindent where $P_{rot}$ is in days and  the typical residuals of the fit are smaller than 17 per cent of the measured periods for a given level of activity. Parameters $A$ and $B$ are listed in Table~\ref{tab:rhk_period_tab}. The final parameters and errors are calculated by bootstrapping simulations. The values of each measurement are randomly moved inside their error bars, calculating a fit for each simulation. The parameters are the median values obtained after 10~000 simulations, and the errors in the parameters the standard deviation of those values. This relationship provides an estimate of the rotation period of stars with low levels of chromospheric activity (up to $\log_{10}R'_{HK} \leq -4.1$) that is compatible with the results of ~\citet{Masca2016} for low-activity GKM stars. For stars with $\log_{10}R'_{HK} \sim -4.0$ $\pm$ 0.1 we support the \citet{Astudillo2017} interpretation of a saturation limit, but unfortunately we do not have any new measurement in that region that could allow for a better constraint.  We did not find any dependence with the spectral subtypes. The estimated parameters for the full sample, early M-dwarfs (M0 to M3) and mid M-dwarfs (M3.5 to M6) are 1$\sigma$ compatible, although the rather small number of mid M-dwarfs studied makes it difficult to assess wether this is a consistent behaviour or not. 

\begin {table*}
\begin{center}
\caption {Parameters for Eq.~\ref{eq_rhk_period} for the different subsets of stars. \label{tab:rhk_period_tab}}
    \begin{tabular}{  l l  c c } \hline
Dataset    & N &  A & B \\ 
\hline
M0 - M6  ($\log_{10}R'_{HK} \leq -4.1$)  & 80 & -- 2.37 $\pm$ 0.28  &  --0.777 $\pm$ 0.054 \\ \\
M0 - M3 ($\log_{10}R'_{HK} \leq -4.1$)  & 66 & -- 2.15 $\pm$ 0.27  &  --0.731 $\pm$ 0.055  \\
M3.5 - M6  ($\log_{10}R'_{HK} \leq -4.1$)  & 14 & -- 2.11 $\pm$ 0.95  &  --0.746 $\pm$ 0.175  \\
 \hline
\end{tabular}  
\end{center}
\end {table*}

Using Eq. ~\ref{eq_rhk_period} with the parameters shown in Table~\ref{tab:rhk_period_tab} it is possible to provide an expected rotation period for a star for which we have a measurement of the $\log_{10}R'_{HK}$. Table~\ref{tab:estimated} shows the estimated rotation period for all the stars for which we could not measure the rotation period. 

\subsection{Rotation-cycle relation}

\begin{figure}[!h]
\includegraphics[width=9cm]{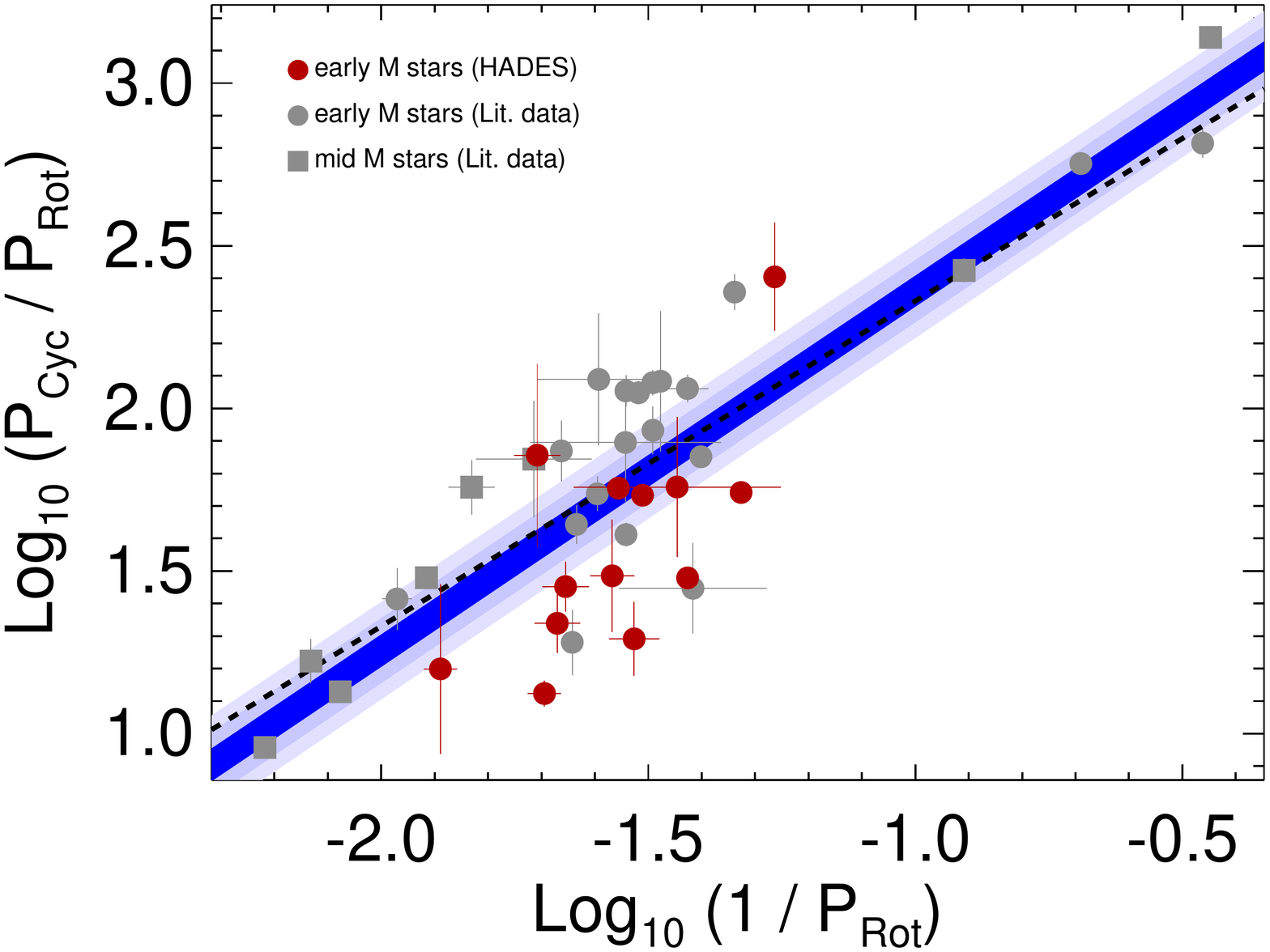}
\caption{$log_{10}(P_{cyc}/P_{rot})$ versus $log_{10}(1/P_{rot})$. The shaded region, going from darker to lighter blue, shows the 1,2,3 $\sigma$ confidence intervals of the best fit. The dashed black line shows the 1-1 relationship. Early M stars refer to M0 to M3 stars, mid M stars are M 3.5 to M6 stars.} 
\label{rot_cyc}
\end{figure}

Solar type stars are known to show a correlation between their rotation period and the length of the magnetic cycle \citep{Baliunas1996, Olah2009, Masca2016}. However this correlation seems to vanish when we move to lower mass stars \citep{Savanov2012, Masca2016}. \citet{Baliunas1996} suggested $P_{cyc}/P_{rot}$ as an observable to study how both quantities relate to each other. It was suggested that the length of the cycle scales as $\sim D^l$, where $l$ is the slope of the relation and $D$ is the dynamo number. Slopes that are different from approximately one would imply a correlation between the length of the cycle and rotation period. Using our new measurements we find a weak correlation between the rotation periods and the cycle lengths. Figure~\ref{rot_cyc} shows the measurements of $P_{cyc}/P_{rot}$ versus 1/$P_{rot}$ of M-dwarfs in a log-log scale. The slope of our current dataset is 1.10 $\pm$ 0.03, which would imply a very weak correlation between both quantities. The detected correlation is very weak, and can be affected by strong biases because of the limited sample. The number of M-dwarfs with measurements of both cycle period and rotation period is still small in comparison with that of G- and K-type stars. We cannot rule out the possibility of having a mix of different types of cycles -- some stars show simultaneous cycles at different timescales -- which would most likely decrease any possible level of correlation. If we measure directly the Spearman's correlation coefficient between the rotation periods and the cycle lengths we obtain a correlation coefficient of -- 0.14 with a 37\% probability of being spurious.  

\subsection{Rotation induced radial velocity signals}

\begin{figure}[!h]
\includegraphics[width=9cm]{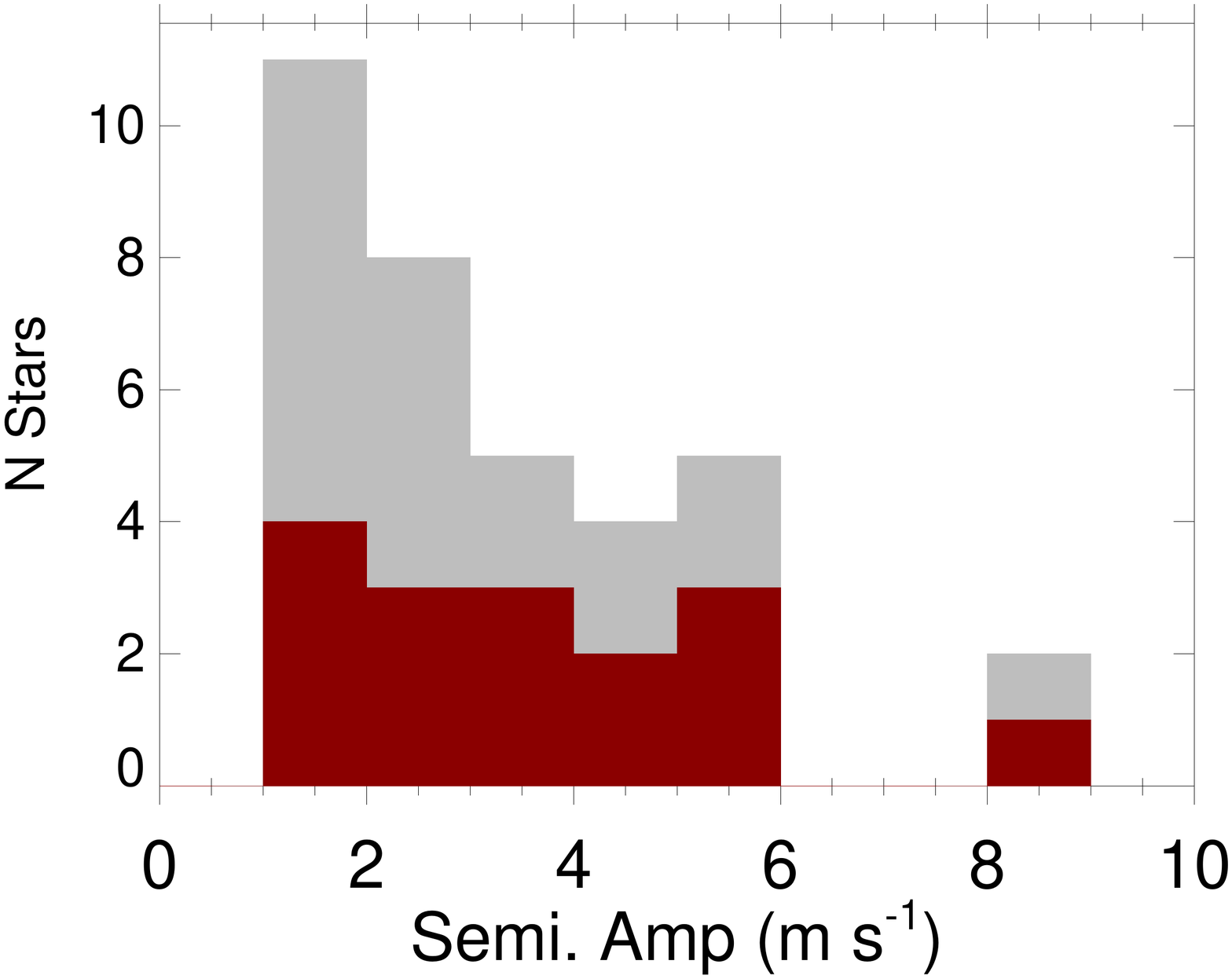}
\includegraphics[width=9cm]{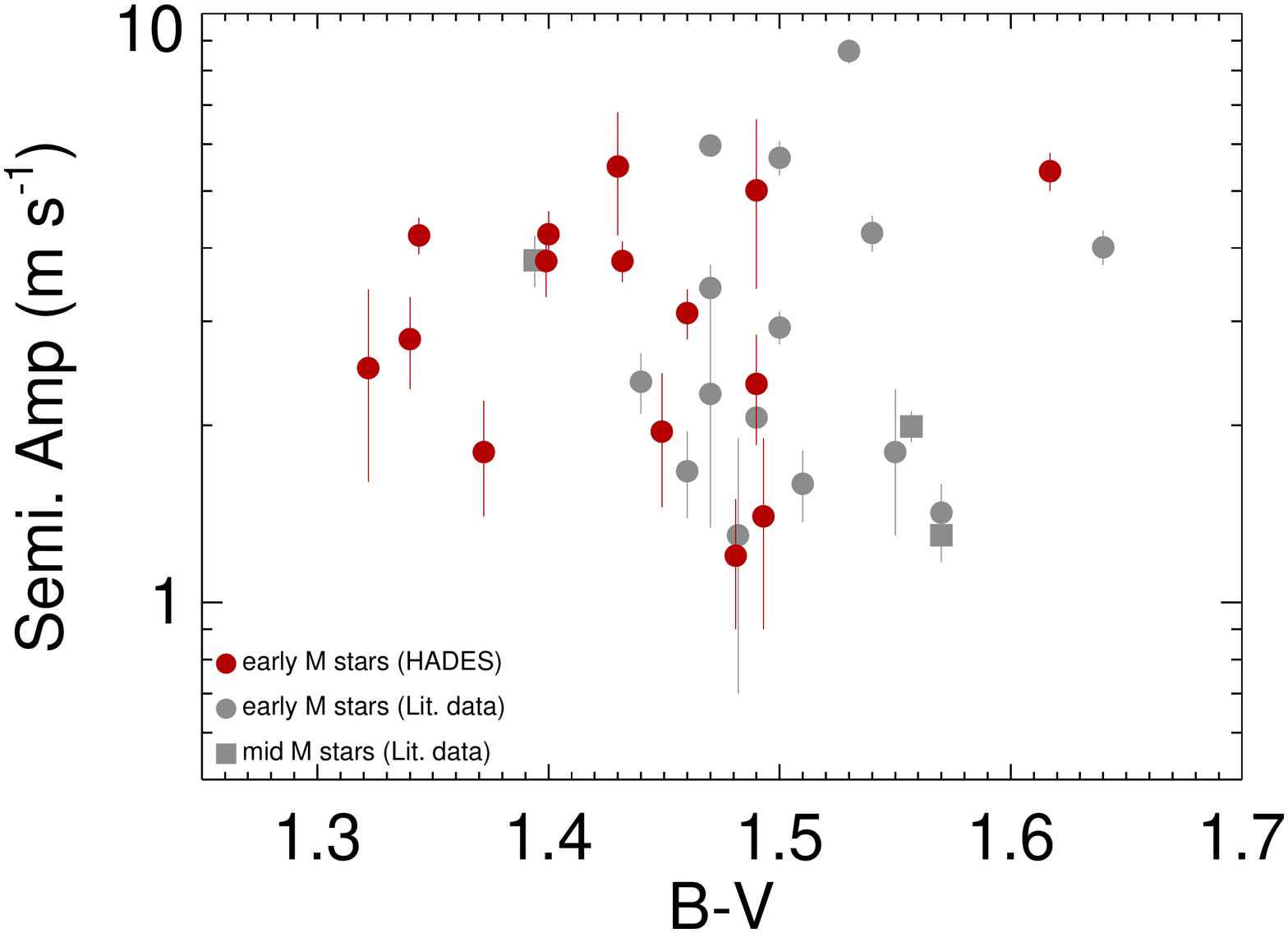}
\caption{\textbf{Top panel:} Distribution of the semi-amplitudes of rotation induced RV signals the for the stars in our sample. The grey area shows the full sample, while the red area highlights the fraction of new measurements. \textbf{Bottom panel:} semi-amplitudes of rotation induced RV signals against the color B-V of the stars. Early M stars refer to M0 to M3 stars, mid M stars are M 3.5 to M6 stars. The Sun value is included as a reference point.}
\label{bv_amp}
\end{figure}

Our dataset includes 35 measurements of rotation induced radial velocity signals of different amplitudes for M-dwarfs. Figure~\ref{bv_amp} shows the distribution of semi-amplitudes of the measured signals and Table~\ref{tab:rv_amp} shows the value of the typical measurements. We find that most signals have very low amplitudes, less than 3 m s$^{-1}$, with only two measurements above 6 m s$^{-1}$, in good agreement with the HADES statistics given in \citet{Perger2017}. Unlike the behaviour of the mean activity level and the rotation period, there does not seem to be any correlation with stellar mass. The typical measured semi-amplitude for the rotation induced signals is very similar to the semi-amplitudes of the Keplerian signals induced by low-mass planets.

\begin {table}
\begin{center}
\caption {Typical semi-amplitude of the rotation induced RV signals in the M-dwarfs under study.\label{tab:rv_amp}}
    \begin{tabular}{  l l  l  l l } \hline
Dataset  & N &  Mean  & Median & $\sigma$  \\ 
        &       &  (m s$^{-1}$) & (m s$^{-1}$) & (m s$^{-1}$) \\\hline
M0 - M6 & 35    & 4.3 & 2.9 &  6.6 \\ 
M0 - M3         & 32 & 4.4 &  2.9 & 6.8 \\
M3.5 - M6       & 3 & 2.4 &  2.0 & 1.3 \\
\\
        
M0 - M3 (HADES) & 16 & 5.6 & 3.8 & 9.5 \\

 \hline
\end{tabular}  
\end{center}
\end {table}

\begin{figure}[!h]
\includegraphics[width=9cm]{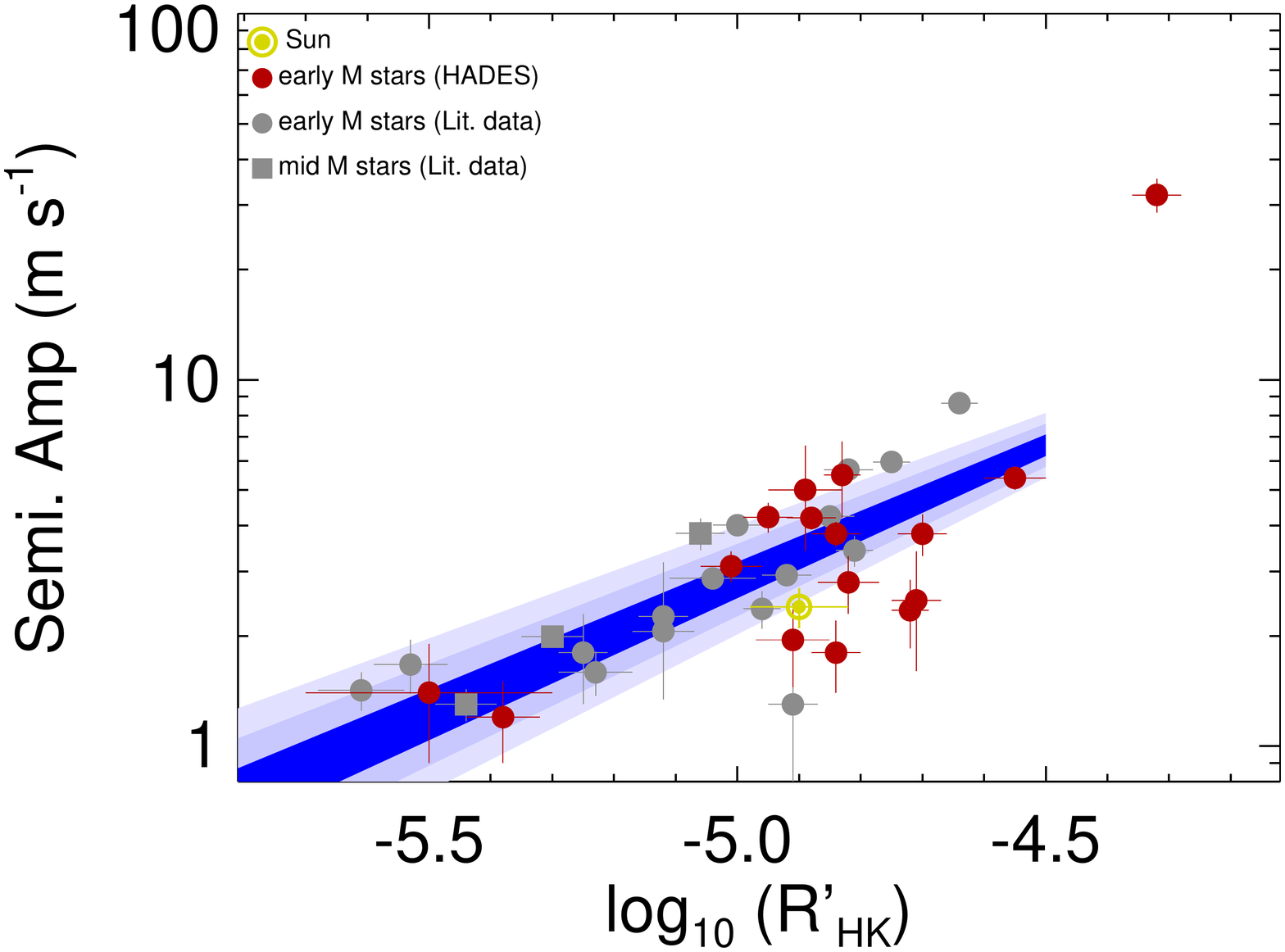}
\caption{RV semi-amplitude of the rotation induced signals vs. chromospheric activity level $\log_{10}R'_{HK}$. The shaded region, going from darker to lighter blue, shows the 1,2,3 $\sigma$ confidence intervals of Eq.~\ref{eq_rhk2} using the parameters of Table~\ref{tab:rhk_amp}. Early M stars refer to M0 to M3 stars, mid M stars are M 3.5 to M6 stars. Sun measurement taken from \citet{Haywood2016}}
\label{rhk_amp}
\end{figure}

\citet{Masca2017b} proposed two different log-linear relationships between the mean chromospheric activity level measured as the $\log_{10}R'_{HK}$ and the mean semi-amplitude of the rotation induced RV signal, one for FGK-type stars and another for M-dwarfs. As expected, the amplitude of the RV signals induced by rotation increases with increased levels of chromospheric activity. The two quantities exhibit a Spearman's correlation coefficient of 0.70 with a probability of being spurious smaller than 10$^{-5}$. Figure~\ref{rhk_amp} shows measured semi-amplitudes against the mean $\log_{10}R'_{HK}$. The new data follows the same trend previously detected. It can be expressed as: 

\begin{equation}
   \log_{10}(K)= A + B \cdot \log_{10}R'_{HK}, 
   \label{eq_rhk2}
\end{equation}

Where K is the semi-amplitude in m s$^{-1}$, and $A$ and $B$ are the parameters shown in Table~\ref{tab:rhk_amp}. The final parameters and errors are calculated by bootstrapping simulations.

\begin {table*}
\begin{center}
\caption {Parameters for Equation~\ref{eq_rhk2} \label{tab:rhk_amp}}
    \begin{tabular}{ l  l  c c c } \hline
 & N & A & B & $\sigma$ Semi. Amp (\%) \\ \hline
 M-type stars ($\log_{10}R'_{HK} \leq -4.5$) & 35 & 4.26 $\pm$ 0.60 & 0.762 $\pm$ 0.061 & 22 \\ \hline
\end{tabular}  
\end{center}
\end {table*}

The measured slope of the RV semi-amplitude with the $\log_{10}R'_{HK}$ is slightly smaller than in \citet{Masca2017b}. Figure~\ref{rhk_amp} shows that the measurements of some of the moderately active stars analysed in the present work show a smaller semi-amplitude than the stars with similar levels of mean chromospheric activity measured in \citet{Masca2017b}, reducing the slope of the relationship. There might also be some effect related to the inclination of the stellar axis with respect to the line of sight. In some of the stars we did find signals at the harmonics of the rotation period which are presumably related to the stellar rotation. It is unclear what semi-amplitudes should be expected in those situations. We caution the reader that the number of stars under study is still relatively low and that additional measurements are required to better constrain the measured parameters. 

There is one outlier to our relationship, the star TYC 2703-706-1 (see Fig.~\ref{rhk_amp}). This star shows the highest level of chromospheric activity as well as the largest induced RV semi-amplitude. It departs from the relationship followed by more quiet stars. At this stage it implies that our relationship can only be applied to stars with $\log_{10}R'_{HK} \leq -4.5$. 

Using Eq. ~\ref{eq_rhk2} with the parameters shown in Table~\ref{tab:rhk_amp} it is possible to estimate an upper limit of the expected semi-amplitude of the rotation induced signal for a star for which we have a measurement of the $\log_{10}R'_{HK}$. Table~\ref{tab:estimated} shows the estimated rotation period and estimated semi-amplitude of the rotation induced signal for all the stars from the HADES RV programme for which we could not detect the stellar rotation and Table~\ref{tab:estimated2} shows the estimated semi-amplitude of the rotation induced signal for those stars where the rotation period could be measured, but we did not detect an induced RV signal. 

\subsection{Inclination of the star's pole to the line of sight}

Using the radii and $v$ sin $i$ measurements from \citet{Maldonado2017} combined with our rotation period determinations it is possible to estimate the inclination of the star's pole to the line of sight. Even for those stars where the period could not be determined from the time series, we have an estimation based on the mean level of chromospheric activity (Table~\ref{tab:estimated}) in a similar way as in \citet{Herrero2012}. The parameters and uncertainties are calculated by bootstrapping simulations. In the situations where the required sin $i$ value  would be bigger than one, a value of one was adopted. In those situations where the $v$ sin $i$ measurement is an upper limit we made the simulations run over all the possible range of $v$ sin $i$. Table~\ref{tab:incli} shows the estimations for the inclination of the star's pole to the line of sight for all the stars in our sample.

For most of the stars the inclination angles are compatible with 90º, but poorly constrained. Even with the large uncertainties in the inclination angles we see that some stars are good candidates for transit searches. The inclinations for stars GJ 625, TYC 2710-691-9, GJ 686, GJ 4196, GJ 2128 and GJ 671 are estimated to be $\sim$  90º. The stars GJ 3942, GJ 548A, GJ 15A, GJ 162, GJ 156.1A, V* BRPsc, GJ 47, TYC 3720-426-1, TYC 3379-1077-1, GJ 9793, NLTT 10614, GJ 408 would also be reasonable candidates, showing inclinations $\textgreater$ 45º. On the other hand the stars TYC 2703-706-1, GJ 49, NLTT 21156, TYC 743-1836-1 and SkTKM 1-650 show estimates of the inclination angle that are not compatible with 90º. In the case of GJ 3998, a search for transit with Spitzer resulted in no evidence (\textit{Affer, private communication}).

Having an estimation of the inclination we can estimate the range of masses of the planets orbiting the stars GJ 3998 \citep{Affer2016}, GJ 625 \citep{Masca2017c}, GJ 3942 \citep{Perger2017b}, and GJ 15A \citep{Howard2014}, assuming their orbits are coplanar with the stellar rotation. We estimate that the masses of GJ 3998 b and c are in the range 2.5 - 4.9 M$_{\oplus}$ and 6.3 - 12.5 M$_{\oplus}$ respectively. For GJ 625 b we obtain a true mass of 2.82 M$_{\oplus}$. For GJ 3942 b we estimate a mass in the range 7.1 - 10.0 M$_{\oplus}$. For the case of GJ 15A b we obtain a range of masses of 5.4 - 5.7 M$_{\oplus}$ if we use the mass provided in \citet{Howard2014}. Recent observations with CARMENES and HARPS-N have added new insight on this system, challenging (\citet{Trifonov2017}) and supporting (Pinamonti et al. {submitted}) the existence of the planet. If we use the mass determined by Pinamonti et al. {submitted}, using the original \citet{Howard2014} data combined with our HARPS-N observations, the mass of GJ 15 A b would be 3.1 - 3.3 M$_{\oplus}$.

\subsection{Filling factor}

Following the approach of \citet{Saar1997} and \citet{Mayor2009} we were able to use our measured and estimated rotation periods and semi-amplitudes of the rotation induced signals to estimate the filling factor in our stars, under the assumption that the radial velocity signal is spot-induced. Assuming the relationship between the measured radial velocity signal and the required filling factor to cause it as:

\begin{equation}
\begin{split}
  K \backsim 6.5 \times f^{0.9} \times v ~\rm{sin} ~i
\end{split}
\label{filling}
\end{equation}

with $K$ being the semi-amplitude of the activity induced radial velocity signal of an equatorial spot in m s$^{-1}$, with a surface filling factor in \% and the $v$ sin $i$ given in km s$^{-1}$. Figure ~\ref{filling} shows the results obtained for our sample. We decided to include both the stars where the parameters have been directly measured, and stars where the parameters have been estimated. Our sample shows a typical filling factor of 0.56 $\pm$ 0.29 \%. The highest measured filling factor is 1.58 \% (1.55 \% for a star in the HADES sample, TYC 2703-706-1), and the minimum is 0.20 \% (0.24 for a star in the HADES sample, GJ 21). We see  the lower envelope of the filling factor increasing towards redder stars and lower masses. The trend of higher filling factors for lower activity (or slowly rotating) stars seems counter-intuitive, but it can be explained as smaller stars with longer rotation periods exhibit a much lower $v ~\rm{sin} ~i$, needing much larger filling factors to cause the same RV semi-amplitude. It has to be noted that for those slowly-rotating stars we are probably biased towards detecting the rotation signals in the stars with the largest filling factors. Those stars with very very small filling factors and very low $v ~\rm{sin} ~i$ are probably below our sensitivity level. For stars where the RV semi-amplitude has been estimated rather than measured it is probably safe to treat our estimation as an upper limit of the expected filling factor.

\begin{figure*}[!h]
\includegraphics[width=18cm]{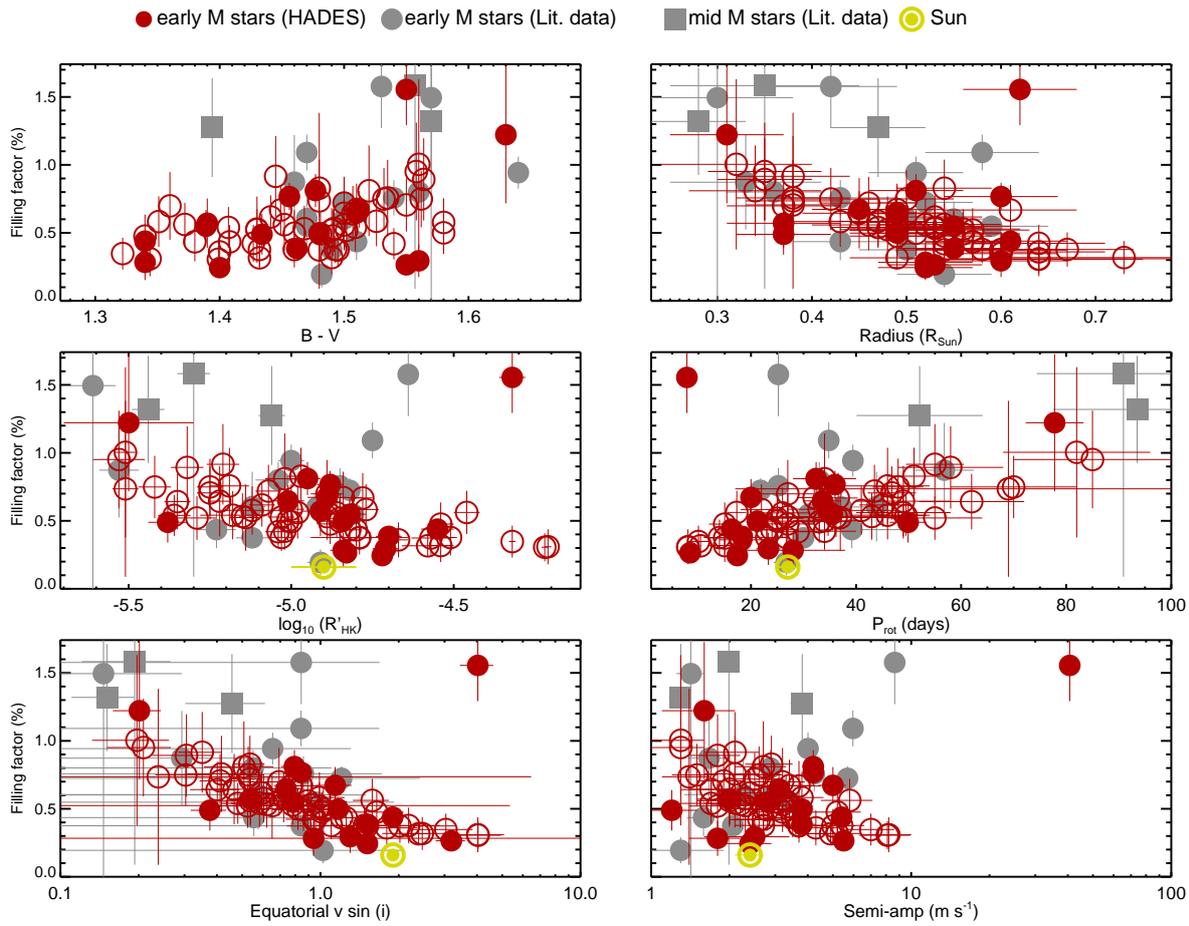}
\caption{Filling factor as a function of color B-V (top left panel), stellar radii (top right panel), $\log_{10}R'_{HK}$ (bottom left panel) and rotation period (bottom right panel). Early M stars refer to M0 to M3 stars, mid M stars are M 3.5 to M6 stars. Empty symbols show stars where the parameters where estimated instead of directly determined. The Sun value is included as a reference point.} 
\label{filling}
\end{figure*}

It is important to re-state that this estimation of the filling factor assumes the RV variations to be caused by spots, which is not always necessarily the case. During the last years there have been a few detections of activity-induced RV variations that are observed in spectroscopic indicators but whose associated photometric signal has not been detected \citep{Kurster2003, Robertson2014, Robertson2015}.This behaviour has been attributed to localized regions on the stellar surface where magnetic fields inhibit convection, but do not create dark spots \citep{Kurster2003}.Some of the stars under study in this work seem to be in the same situation. We could only detect the photometric modulation in four of the 18 stars with sufficient photometric data to allow for a detection. In particular for the stars GJ 4057, GJ 3997, GJ 4306, GJ 3998, GJ 49, V* BRPsc, and GJ 552 we are able to detect the rotation-induced signal in at least one activity indicator, but not in the photometric data. Future studies with higher precision photometry would be needed to confirm whether or not the RV signals in these stars are caused by dark spots or not, and to confirm whether or not our filling-factor estimations are realistic.

\section{Conclusions}

We report an analysis of the magnetic activity of 71 early M-dwarfs using HARPS-N spectra, obtained as part of the HADES RV survey, with HARPS spectra and ASAS light curves used as supplementary data. We report 72 new $\log_{10}R'_{HK}$ measurements, 33 new rotation period measurements, 18 new magnetic cycle detections, 16 new rotation induced RV signals and 36 estimations of the rotation period and the amplitude of the rotation induced RV signals using activity-rotation relationships. 

We measure the typical mean level of chromospheric activity, measured as the $\log_{10}R'_{HK}$, for the M-dwarfs in our sample at $\log_{10}R'_{HK}$ $\sim$ --5. The star-by-star variance however is much larger than for the case of solar type stars, going from very high levels, with $\log_{10}R'_{HK}$ $\sim$ --4 to extremely low levels at $\log_{10}R'_{HK}$ $\sim$ --6.0. There seems to be a negative slope of the lower envelope towards lower mass stars, meaning lower mass stars in our sample show lower mean levels of chromospheric activity. 

We find that the mean rotation period for the M-dwarfs in our sample is $\sim$ 40 days with a very large variance of $\sim$ 34 days. Most M-dwarfs in our sample are very slow rotators, with periods longer than 20 days, reaching periods longer than 150 days in some cases. The distribution of rotation periods shows two dominant peaks, at periods shorter than five days and at periods around 40 days, which probably correspond to two different age groups. Similar to what happens with the distribution of the $\log_{10}R'_{HK}$ values, there is a slope towards lower mass stars. Low mass stars in our sample are slower rotators than more massive stars. The length of the magnetic cycles detected around M-dwarfs ranges at least from  less than two years to over 14 years, with a mean length of 6.5 years. 

We confirm and refine the activity-rotation relation proposed in \citet{Masca2015, Masca2016} for M-type stars.  We also confirm and refine the relation proposed in \citet{Masca2017b} that estimates the amplitude of the rotation induced RV signal as a function of the mean chromospheric activity level. 

Using previous $v$ sin $i$ measurements we estimate the inclinations of the star's poles to the line of sight for all the stars in the sample, which identifies some of them as promising candidates for transit searches. We also estimate the range of true masses of the planets GJ 3998 b and c (2.5 - 4.9 M$_{\oplus}$ and 6.3 - 12.5 M$_{\oplus}$), of the planet GJ 625 b (2.82 M$_{\oplus}$), of GJ 3942 b (7.1 - 10.0 M$_{\oplus}$), and of GJ 15A b (5.4 - 5.7 M$_{\oplus}$).

\section*{Acknowledgements}
A.S.M acknowledges financial support from the  Swiss National Science Foundation (SNSF). This work has been financed by the Spanish Ministry project MINECO AYA2014-56359-P. J.I.G.H. acknowledges financial support from the Spanish MINECO under the 2013 Ram\'on y Cajal program MINECO RYC-2013-14875.  M. P., I. R., J.C.M., E.H., A. R., and M.L. acknowledge support from the Spanish Ministry of Economy and Competitiveness (MINECO) and the Fondo Europeo de Desarrollo Regional (FEDER) through grant ESP2016-80435-C2-1-R. GAPS acknowledges support from INAF through the PREMIALE WOW 2013 research project. L. A. and G. M. acknowledge support from the Ariel ASI-INAF agreement N. 2015-038-R.0. The research leading to these results has received funding from the European Union Seventh Framework Programme (FP7/ 2007-2013) under Grant Agreement No. 313014 (ETAEARTH). G.S.\ acknowledges financial support from \lq\lq\ Accordo ASI--INAF\rq\rq\ n. 2013-016-R.0 July 9, 2013. This research has made extensive use of the SIMBAD database, operated at CDS, Strasbourg, France and NASA's Astrophysics Data System.

%
%
\bibliography{phot_ref}

\onecolumn
\begin{appendix}
\section{Tables: Data and results} \label{append_a}

\begin{longtable}{llccccccccll@{}}
\caption {Available data for all the stars under study. Stars are sorted by spectral subtype, and within a given spectral subtype by B-V color. \label{tab:star_parameters}}\\
\endfirsthead
\multicolumn{8}{c}{\tablename\ \thetable\ -- \textit{Continued from previous page}} \\
\hline
\endhead
\hline \multicolumn{8}{r}{\textit{Continued on next page}} \\
\endfoot
\multicolumn{10}{l}{$^{*}$ 2MASS J22353504+3712131}  \\
\multicolumn{10}{l}{\textbf{Reference for spectral types:} \citet{Maldonado2017}} \\
\multicolumn{10}{l}{\textbf{References for magnitudes:} 1 - \citet{Zacharias2013}, 2 - \citet{Hog2000}, 3 - \citet{Koen2010}, 4 - \citet{Epps1972},}\\
\multicolumn{10}{l}{5 - \citet{Mumford1956}, 6 - \citet{Karchenko2001}, 7 - \citet{Oja1991}, 8 - \citet{Mermilliod1986}, 9 - \citet{MonetLevine2003}, 10 - \citet{Landolt2009},} \\
\multicolumn{10}{l}{11 - \citet{Weistrop1981}} \\
\endlastfoot
\hline

Star & Sptype & B & V &  N$_{obs ~\rm HARPS-N}$ & N$_{obs ~\rm HARPS-S}$ & T$_{\rm Span ~Spec. ~(yr)}$ &N$_{obs ~\rm ASAS}$ & T$_{\rm Span ~Phot. ~(yr)}$ & Ref.\\ \hline
TYC 2710-691-1         &   K7.5  & 13.14 & 11.52 & 1 &&& & &1\\
2MASS J2235 $^{*}$            &   K7.5  &       & 11.77 & 4 & & 0.5  & & & 1\\
GJ 3942                                 &       M0      & 11.59 & 10.25  & 135   & & 3.3 &  & & 2\\
GJ 4057                                 &       M0              & 12.26 & 10.77    & 109  & & 4.3 & 318 & 6.6 & 3 \\
TYC 1795-941-1         &   M0       & 12.20 & 11.06 & 2 & & 2.8 & &  & 1\\
TYC 3720-426-1         &   M0 & 12.05  &  11.25 & 3 & & 0.2 &&  & 2,1\\
TYC 3379-1077-1        &   M0 & 12.86 &  11.43 & 11 & & 3.8 & &  & 1\\
TYC 743-1836-1         &   M0 & 12.03  & 10.71 & 4 & & 2.8 & 325 & 7.0 &  2,1\\
GJ 548A                &   M0 & 11.169 & 9.72 & 34 & & 2.5 & &  & 3\\
GJ 3997                &   M0 & 12.20   & 10.59  & 91   & & 4.3 & 445 & 6.6 & 1 \\
GJ 731                 &   M0 & 11.585 &  10.10 & 34 & & 3.3 & 411 & 6.6 & 3\\
BPM 96441              &   M0 & 13.3  &   11.9 & 16 & & 3.5 & & & 1 \\
NLTT 53166             &   M0 & 12.5   & 11.01 & 26 & & 3.5 & & & 2\\
GJ 9793                &   M0 & 12.10   & 10.51  & 26   & & 3.4 & 462 & 6.5 & 2\\
GJ 184                                  &       M0.5            & 11.36 & 9.94     & 63   & & 3.9 & & & 1\\
GJ 694.2                                &       M0.5            & 12.09 & 10.86    & 129  & & 3.3 & & & 2,1\\
GJ 9689                                 &   M0.5                & 12.67 & 11.23    & 99   & & 3.5 & 273 & 6.5 & 2,1\\
GJ 740                                  &   M0.5        & 10.67 & 8.29   & 93    & 53 & 8.3 &  458 & 8.1 & 3\\
GJ 685                                  &   M0.5        & 11.45 & 9.97   & 56     & & 3.3 & & & 4\\
GJ 720A                     &   M0.5    & 11.26 & 9.98   & 84   & & 3.2 & & & 1\\
GJ 3822                                 &   M0.5        & 12.10 & 10.64  & 63    & 57 & 8.9 & 312 & 6.5 & 3\\
GJ 1074                                 &   M0.5        & 12.39 & 10.93  & 28    & & 3.8 & & & 5 \\
GJ 3352                &   M0.5    & 12.56 &  11.07 & 12 & & 4.1 & & & 6\\
NLTT 4188              &   M0.5    & 12.97 &  11.36 & 1 & & && &  2,1\\
StKM 1-650            &   M0.5    & 13.36 & 11.87 & 13 & & 3.8 & & & 1,6\\
GJ 412A                         &       M0.5            & 10.27 & 8.78   & 96    & & 3.7 & & & 7\\
GJ 9404                                 &   M0.5                & 12.25 & 10.64    & 52   & & 3.8 &  77 & 1.5 & 2,1\\
TYC 2703-706-1                  &   M0.5                & 13.24 & 11.69  & 68    & & 3.5 & & & 2 \\
GJ 4092                &   M0.5        & 12.24 & 10.86 & 26 & 11 &8.2 &386 & 8.1 & 2,1\\
GJ 4306                                 &       M1      & 11.76 & 10.47  & 135   & &3.6 & 241 & 6.4 & 2,1\\
GJ 21                                   &   M1  & 12.03 & 10.56  & 104          & & 4.0 & &  & 1\\
GJ 150.1B                               &   M1          & 12.27 & 10.76  & 63    & & 4.2 &348 & 7.0& 3\\
GJ 3998                                 &   M1          & 12.34 & 10.83  & 172   & 6 & 8.4 & 542 & 7.1 & 3\\
GJ 2                                    &   M1          & 9.97  & 9.2    & 96    & & 3.7 & & & 8  \\
GJ 15A                                  &   M1          & 9.63  & 8.13   & 113   & & 3.5 & & & 8\\
GJ 3186                &   M1  & 12.5  &  11.05 & 1 & &&  117 & 7.0 & 1,6\\
GJ 119A                &   M1 & 11.77   & 10.46  & 106 & & 3.2 & &  & 2 \\
GJ 162                                  &   M1  & 11.69 & 10.16  & 66    & & 3.9 & & & 1\\
GJ 272                              &   M1 & 11.69  & 10.213 & 6 & & 0.8 & & & 2\\
GJ 686                 &   M1 & 11.11 & 9.58 & 11 & 21 & 12.2 & & & 3\\
GJ 4196                &   M1 & 12.69  & 11.11 & 1 & & & & & 2\\
GJ 606                                  &   M1.5                & 12.00 & 10.49    & 27   & 24 & 9.1 & 425 & 6.5 & 3\\
GJ 49                                   &   M1.5        & 11.06 & 9.5    & 116   &  & 4.0 & &  & 1,2\\
GJ 156.1A                       &   M1.5        & 12.29 & 10.86  & 81   & & 4.0 & & & 1 \\
GJ 16                                   &   M1.5        & 12.39 & 10.86  & 107   & 7 &8.9 & 302 & 8.8& 3 \\ 
GJ 3014                &   M1.5  & 12.80  & 11.12 & 1 & & & & & 2\\
NLTT 10614             &   M1.5  & 12.8   & 11.31 & 1 &  & & & & 9,1\\
GJ 3649                &   M1.5  & 12.19  & 10.85 & 22 & & 3.6 & & & 2\\
GJ 9440                &   M1.5 & 12.16 & 10.55  & 98   & & 3.9 & & & 1\\
GJ 521A                &   M1.5 & 11.50 & 10.26  & 97   & & 3.9 & & & 2,1\\GJ 895                 &   M1.5 & 11.66  &  10.11 & 8 & & 4.0 & & & 1\\
V* BRPsc               &   M1.5 & 10.43 & 8.99   & 10   & 68     & 10.6 & 307 & 9.0 &  10\\
GJ 450                 &   M1.5 & 11.30 & 9.80 & 34 & & 3.7 & & & 1\\           
GJ 625                                  & M2            & 11.80 & 10.17  & 140   & & 3.4 & &  & 1 \\
GJ 414B                         & M2            & 12.39 & 10.72  & 34   & & 2.8 & &  & 2,1 \\
GJ 47                                   & M2            & 12.26 & 10.87  & 77    & &  3.3 & &  & 6 \\    
NLTT 21156                              & M2            & 12.7  & 11.2   & 40    & & 3.8 & &  & 1,9 \\
GJ 1030                & M2 & 12.91 & 11.40 & 4 & 8 & 3.0 & 225 & 6.9 & 3\\
GJ 552                                  & M2    & 12.18 & 10.68  & 58    & 8 & 8.9 &  323 & 6.5 & 3\\
NLTT 52021             & M2 & 12.61  & 11.34 & 1 & & & & & 2,1\\
GJ 70                                   & M2.5          & 12.45 & 10.92  & 23    & 9 & 9.2 & 312 & 8.9 & 3\\
GJ 2128                                 & M2.2          & 13.09 & 11.53  & 25    & 6     & 8.3 & 458 & 6.6 & 3\\
GJ 399                                  & M2.5  & 12.79 & 11.28  & 33    & 8      & 9.0 & 629 & 7.0 & 3\\
GJ 408                                  & M2.5          & 11.58 & 10.02  & 45     & &3.6 &  279 & 6.5 & 3\\
GJ 671 & M2.5 & 12.93 &11.370 & 6 & &0.7 & &  & 1\\
GJ 3117A & M2.5 & 12.73 &  11.38 & 12 & & 3.3 & &  & 2,1\\
GJ 26    & M2.5 & 12.60  & 11.24 & 52 & & 3.4 & & & 1\\
GJ 476                                  & M3    & 12.85 & 11.41  & 14   & 12      & 8.7   & 372 & 8.0 & 3\\
GJ 793                                  & M3            & 12.00 & 11.68  & 46 & & 2.4 & &  & 2   \\
GJ 3126                & M3 & 12.45  & 10.95 & 6 & & 3.3 & &  & 2\\
GJ 119B                & M3 &13.07  &  11.65 & 7 & & 3.3 & & & 11\\
 \hline
\end {longtable}

\newpage
    
\begin{longtable}{llcccccccc@{}}
\caption {Measured cycle lengths, rotation periods and semi-amplitude of the rotation induced RV signals. Results are the weighted mean of the individual detections of the different activity proxies time series. Errors are the standard deviation of the individual measurements. The column 'Det' identifies the activity proxies used for each individual star. \label{tab:results}}\\
\endfirsthead
\multicolumn{10}{c}{\tablename\ \thetable\ -- \textit{Continued from previous page}} \\
\hline
\endhead
\hline \multicolumn{10}{r}{\textit{Continued on next page}} \\
\endfoot
\hline 
\multicolumn{8}{l}{\textbf{Detection:} 1 - S$_{MW}$ index, 2 - H$\alpha$, 3 - FWHM, 4 - RV, 5 - ASAS  m$_{V}$ light-curve} \\
\multicolumn{8}{l}{$^{a}$ detected using EXORAP photometry in \citet{Perger2017b}} \\
\multicolumn{8}{l}{$^{b}$ detected using Fairborn Observatory photometry in \citet{Howard2014}} \\
\multicolumn{8}{l}{$^{c}$ detected using EXORAP photometry in \citet{Masca2017c}} \\
\endlastfoot
\hline    

Star & $\log_{10}R'_{HK}$ &   $P_{cycle}$  & $FAP $& Det &  $P_{rot}$  &$FAP $  &Det  & $S. Amp. ~P_{rot} $ & $S. Amp. ~P_{rot}/2 $\\ 
        &       &  $(yr)$       & $(\%)$        & &$(d)$        &       $(\%)$  & & $(m s^{-1} )$ & $(m s^{-1} )$\\\hline

GJ 3942                                 &       -- 4.55 $\pm$   0.05 & &  &  & 16.3  $\pm$  0.1 & < 0.1 & 1,2,4$^{a}$ & 5.4 $\pm$ 0.4\\
GJ 4057                                 &        -- 5.03 $\pm$ 0.04 & 2.6 $\pm$ 0.6 & < 1 & 1,2& 26.7 $\pm$ 0.1 & < 0.1 & 1,2\\

GJ 548A                &   -- 4.78 $\pm$ 0.03 & & & & 36.6 $\pm$ 0.1 & 1  & 1\\
GJ 3997                &   -- 4.78 $\pm$ 0.03 & 2.2 $\pm$ 1.3 & < 0.1 & 1,2 & 37 $\pm$ 13 & < 0.1 & 1,2\\

GJ 184                                  &       -- 5.09 $\pm$ 0.04      & 3.5 $\pm$ 0.5  & < 0.1 & 2 &  45.0 $\pm$ 0.1 & < 0.1 & 1\\
GJ 694.2                                &       -- 4.46 $\pm$ 0.04 & > 10 & < 0.1 & 1 & 17.3 $\pm$ 0.1 & < 0.1 & 1        \\
GJ 9689                                 &    -- 4.82 $\pm$ 0.05 & > 10 & < 0.1 & 1 & 35.7 $\pm$ 0.2 & < 0.1 & 1,2,4 & 2.8 $\pm$ 0.5 & 2.2 $\pm$ 0.6\\
GJ 740                                  &   -- 4.88 $\pm$ 0.04 & 5.6 $\pm$ 0.1 & < 0.1 & 1,2 & 36.3 $\pm$ 1.7 & < 0.1 & 1,2,3,4,5 &  3.4 $\pm$ 0.3 & 1.8 $\pm$ 0.3\\
GJ 685                                  &   -- 4.79 $\pm$ 0.04 & & & & 16.3 $\pm$ 4.2 & < 0.1 & 1,2,4\\
GJ 720A                     &   -- 5.03 $\pm$ 0.04 & 2.3 $\pm$ 0.5 & < 0.1 & 2 & 34.5 $\pm$ 4.7 & < 0.1 & 1\\
GJ 3822                                 &   -- 4.70 $\pm$ 0.04 & 12.7 $\pm$ 4.1 & < 0.1 & 1,2 & 18.3 $\pm$ 0.1 & < 0.1 & 1,2,4,5& 3.8 $\pm$ 0.5\\
GJ 606                                  &   -- 4.89 $\pm$ 0.06 & & & & 20.0 $\pm$ 2.0 & 1 & 1,2 & 5.0 $\pm$ 0.6 \\
GJ 412A                         &       -- 5.51 $\pm$ 0.04 & && & 100.9  $\pm$ 0.3  & < 1 & 3    \\
GJ 9404                                 &   -- 4.71 $\pm$ 0.04 & & & & 23.2 $\pm$ 0.1  & < 1 & 1,2,4 & 2.5 $\pm$ 0.9 \\
TYC 2703-706-1                  &   -- 4.32 $\pm$ 0.04  & & & &7.8 $\pm$ 0.2 & < 1       & 1,2,4 &  40.7 $\pm$ 3.4 \\
GJ 4306                                 &        -- 4.84 $\pm$ 0.04 & 4.4 $\pm$ 0.9 & < 0.1 & 1,2,3 & 27.0 $\pm$ 2.5 & < 0.1 & 3,4 & 1.8 $\pm$ 0.4\\
GJ 21                                   &   -- 4.72 $\pm$ 0.03  & & & & 17.4 $\pm$ 1.1 & < 0.1 & 1,2,3,4 & 2.4 $\pm$ 0.5 & 4.1 $\pm$ 0.5 \\
GJ 3998                                 &   -- 5.01 $\pm$ 0.05  & 1.8 $\pm$ 0.4 & < 0.1 & 1,2,3 & 33.6 $\pm$ 3.6 & < 0.1 & 1,2,3,4 & 3.1 $\pm$ 0.3\\
GJ 2                                    &   -- 4.84 $\pm$  0.04  & 3.2 $\pm$ 0.1 &  < 0.1  & 1& 21.2 $\pm$  0.5  & < 0.1 & 1,2,4 & 3.8 $\pm$ 0.3 & 2.8 $\pm$ 0.3\\
GJ 15A                                  &   -- 5.27 $\pm$ 0.04  & 2.8 $\pm$ 0.5 & < 0.1 & 1,2,3 &  45.0 $\pm$ 4.4 & < 0.1 & 1,3$^{b}$       \\
GJ 119A                &     -- 4.97 $\pm$ 0.04 & > 10 & < 0.1 & 1 &  51.2 $\pm$ 4.4 & < 0.1& 1,2\\
GJ 162                                  &   -- 4.95 $\pm$ 0.04 & 4.8 $\pm$ 0.2 & < 0.1 & 1 & 32.4 $\pm$ 1.6 & < 0.1 & 1,2,3,4 & 4.2 $\pm$ 0.4       \\
GJ 49                                   &   -- 4.83 $\pm$ 0.03 & > 7 & < 1& 1,2 & 18.4 $\pm$ 0.7 & < 0.1 & 1,2,3,4 & 5.5 $\pm$ 0.3 & 2.2 $\pm$ 0.3  \\
GJ 156.1A                       &   -- 5.04 $\pm$ 0.04  & & & & 41.2 $\pm$ 6.3  & < 1 & 1,2 \\
GJ 9440                             &   -- 5.25 $\pm$ 0.04      & & & & 48.0 $\pm$ 4.8 & < 0.1 & 1,2 \\
GJ 521A                &   -- 5.36 $\pm$ 0.05 & 1.8 $\pm$ 0.1  & < 1 & 1 &  49.5 $\pm$ 3.5 & < 0.1 & 2,3 \\
V* BRPsc               &   -- 5.38 $\pm$ 0.06 & > 10 & < 0.1 & 1& 49.9 $\pm$ 3.5 & 0.1 & 1,2 &   1.2 $\pm$ 0.3\\           
GJ 625                                  &       -- 5.50 $\pm$ 0.20 & 3.3 $\pm$ 1.5 & < 1,2 & 1 & 77.8 $\pm$ 5.5 & < 0.1 & 1,2,3,4$^{c}$ & 1.8 $\pm$ 0.5\\
GJ 47                                   &       -- 4.91 $\pm$ 0.06 & & & &  34.7 $\pm$ 0.1 & < 0.1 & 1,2,4 & 2.0 $\pm$ 0.5 \\    
NLTT 21156                              &       -- 4.58 $\pm$ 0.08 & & & & 10.4 $\pm$ 0.1  & < 1 & 2 \\
GJ 1030                &   -- 4.84 $\pm$  0.06  & & &   & 32.0 $\pm$  3.0       & < 0.1 & 5\\
GJ 552                                  &   -- 5.07 $\pm$ 0.04 & & & & 43.5 $\pm$ 1.5  & < 0.1 & 1 ,2\\
GJ 476                                  &   -- 5.21 $\pm$ 0.05 & & & & 55.0 $\pm$   5.5 & < 1 & 2,5\\ \hline

\end {longtable}

\newpage

\begin{longtable}{l  c c c @{}}
\caption {Expected rotation periods and semi-amplitudes of the rotation induced RV signals for the stars for which a direct determination was not possible \label{tab:estimated}}\\
\endfirsthead
\multicolumn{4}{c}{\tablename\ \thetable\ -- \textit{Continued from previous page}} \\
\hline
\endhead
\hline \multicolumn{4}{r}{\textit{Continued on next page}} \\
\endfoot
\hline \multicolumn{4}{l}{$^{*}$ 2MASS J22353504+3712131} 
\endlastfoot
\hline

Star & $\log_{10}R'_{HK}$ &   Expected $P_{rot}$ &   Expected $~Semi. Amp$\\ 
        &       & $(d)$ & $(m s^{-1})$\\\hline
TYC 2710-691-1         &   -- 5.02 $\pm$ 0.01 & 34 $\pm$ 6 & 2.7 $\pm$ 0.6\\
2MASS J2235$^{*}$             &   -- 4.51 $\pm$ 0.02 & 14 $\pm$ 3 & 5.4 $\pm$ 1.2 \\
TYC 1795-941-1         &   -- 4.21 $\pm$ 0.01 & 8 $\pm$ 1 & \\
TYC 3720-426-1         &   -- 4.22 $\pm$ 0.01 & 8 $\pm$ 1 &\\
TYC 3379-1077-1        &   -- 4.79 $\pm$ 0.15 & 23      $\pm$ 4 & 3.7 $\pm$ 0.9 \\
TYC 743-1836-1         &   -- 4.32 $\pm$ 0.01   & 10 $\pm$ 2 & \\
GJ 731                 &   -- 5.00 $\pm$ 0.04 & 33 $\pm$ 6 & 2.8 $\pm$ 0.6 \\
BPM 96441              &   -- 4.67 $\pm$ 0.03 & 18 $\pm$ 4 & 4.3 $\pm$ 1.0 \\
NLTT 53166             &   -- 5.29 $\pm$ 0.06 & 55 $\pm$ 9 & 1.8 $\pm$ 0.4\\
GJ 9793                &   -- 4.54 $\pm$ 0.03 & 15 $\pm$ 3 & 5.2 $\pm$ 1.2 \\
GJ 1074                                 &   -- 4.85     $\pm$ 0.06 &  25 $\pm$ 5 & 3.4 $\pm$ 0.8\\
GJ 3352                &   -- 4.88 $\pm$ 0.03 & 27 $\pm$ 5 & 3.2 $\pm$ 0.7\\
NLTT 4188              &   -- 4.80 $\pm$ 0.01 & 23 $\pm$ 4 & 3.6 $\pm$ 0.8 \\
SkTKM 1-650            &   -- 4.57 $\pm$ 0.06 & 15 $\pm$ 3 & 5.0 $\pm$ 1.1\\
GJ 4092                                 &       --4.84 $\pm$ 0.04 & 25 $\pm$ 5 & 3.4 $\pm$ 0.8 \\
GJ 150.1B                               &   -- 4.85 $\pm$ 0.04 & 25 $\pm$ 5 & 3.4 $\pm$ 0.8\\
GJ 3186                &   -- 4.98 $\pm$ 0.01 & 49 $\pm$ 9 & 2.0 $\pm$ 0.5\\
GJ 272                              &   -- 5.12 $\pm$ 0.10 & 41 $\pm$ 7 & 2.3 $\pm$ 0.5 \\
GJ 686                 &   -- 5.42 $\pm$ 0.05 & 70 $\pm$ 12 & 1.5 $\pm$ 0.3\\
GJ 4196                &   -- 5.01 $\pm$ 0.01 & 34 $\pm$ 6 & 2.7 $\pm$ 0.6 \\

GJ 16                                   &   -- 5.00 $\pm$ 0.04 & 33 $\pm$ 6 & 2.8 $\pm$ 0.6               \\ 
GJ 3014                &   -- 4.82 $\pm$ 0.01 & 24 $\pm$ 5 & 3.5 $\pm$ 0.8\\
NLTT 10614             &   -- 5.14 $\pm$ 0.08 & 43 $\pm$ 8 & 2.2 $\pm$ 0.5\\
GJ 3649                &   -- 4.54 $\pm$ 0.04 & 15 $\pm$ 3 & 5.2 $\pm$ 1.2\\
GJ 895                 &   -- 4.83 $\pm$ 0.04 & 24 $\pm$ 5 & 3.5 $\pm$ 0.8\\
GJ 414B                         &       -- 5.35 $\pm$ 0.04 & 62 $\pm$ 10 & 1.7 $\pm$ 0.4\\
NLTT 52021             & -- 4.89 $\pm$ 0.01 & 27 $\pm$ 5 & 3.2 $\pm$  0.7\\
GJ 70                                   & -- 5.19 $\pm$ 0.06 & 46 $\pm$ 8 & 2.1 $\pm$ 0.5\\
GJ 2128                                 & -- 5.53 $\pm$ 0.08 & 85 $\pm$ 15 & 1.3 $\pm$ 0.3 \\
GJ 399                                  & -- 5.18 $\pm$ 0.10 & 46 $\pm$ 8 & 2.1 $\pm$ 0.5\\
GJ 408                                  & -- 5.32 $\pm$ 0.05 & 58 $\pm$ 10 & 1.8 $\pm$ 0.4 \\
GJ 671 &                  -- 5.51 $\pm$ 0.03 & 82 $\pm$ 14 & 1.3 $\pm$ 0.3\\

GJ 3117A &               -- 4.77 $\pm$ 0.04 & 22 $\pm$ 4 & 3.8 $\pm$ 0.8 \\
GJ 26    &               -- 4.89 $\pm$ 0.05 & 27 $\pm$ 5 & 3.2 $\pm$  0.7\\

GJ 793                                  & -- 5.02 $\pm$ 0.06 & 34 $\pm$ 6 & 2.7 $\pm$ 0.6 \\
GJ 3126                & -- 5.22 $\pm$ 0.03 & 49 $\pm$ 9 & 2.0 $\pm$ 0.6\\
GJ 119B                & -- 4.86 $\pm$ 0.04 &  26 $\pm$ 5 & 3.3 $\pm$ 0.7 \\
 \hline

\end {longtable}

\newpage
    
\begin{longtable}{l  c c  @{}}
\caption {Expected maximum semi-amplitudes of the rotation induced RV signals for the stars for which a direct determination of the period was possible but the signal did not show up in the RV time series\label{tab:estimated2}}\\
\endfirsthead
\multicolumn{3}{c}{\tablename\ \thetable\ -- \textit{Continued from previous page}} \\
\hline
\endhead
\hline \multicolumn{3}{r}{\textit{Continued on next page}} \\
\endfoot
\hline
\endlastfoot
\hline
    
Star & $\log_{10}R'_{HK}$ &     Expected $~Semi. Amp$\\ 
        &        & $(m s^{-1})$\\\hline

GJ 4057 & -- 5.03 $\pm$ 0.04 & 2.7 $\pm$        0.6\\
GJ 548A & -- 4.78 $\pm$ 0.03 & 3.7 $\pm$ 0.8\\
GJ 3997 & -- 4.78 $\pm$ 0.03 & 2.3 $\pm$ 0.5\\
GJ 184  & -- 5.09 $\pm$ 0.04 & 2.4 $\pm$  0.5\\
GJ 694.2 & -- 4.46 $\pm$ 0.04 & 5.8 $\pm$ 1.3\\
GJ 685 & --4.79 $\pm$ 0.04 & 3.7 $\pm$ 0.8\\
GJ 720A & -- 5.03 $\pm$ 0.04 & 2.7 $\pm$ 0.6\\
GJ 412A & -- 5.51 $\pm$ 0.04 & 1.4 $\pm$ 0.3\\
GJ 15A & -- 5.27 $\pm$ 0.04 & 1.9 $\pm$ 0.4\\
GJ 119 A & -- 4.97 $\pm$ 0.04 & 2.9 $\pm$ 0.6\\
GJ 156.1A & -- 5.04 $\pm$ 0.04 & 2.6 $\pm$ 0.6\\
GJ 9440 & -- 5.25 $\pm$ 0.04 & 2.0 $\pm$ 0.4\\
GJ 521 A & -- 5.36 $\pm$ 0.05 & 1.7 $\pm$ 0.4\\
NLTT 21156 & -- 4.58 $\pm$ 0.08 & 4.9 $\pm$ 1.1\\
GJ 1030 &-- 4.84 $\pm$  0.06 & 3.4 $\pm$   0.8\\
GJ 552 & -- 5.07 $\pm$ 0.04 & 2.5 $\pm$  0.6\\
GJ 476 & -- 5.21 $\pm$ 0.05 & 2.1 $\pm$ 0.5 \\ \hline

\end {longtable}

\newpage

\begin{longtable}{l c | l c  @{}}
\caption {Inclination of the star's pole to the line of sight. Left column shows the stars for which a direct determination of the period was possible. Right column shows the stars where the period was estimated using the activity-rotation relationship. \label{tab:incli}}\\
\endfirsthead
\multicolumn{4}{c}{\tablename\ \thetable\ -- \textit{Continued from previous page}} \\
\hline
\endhead
\hline \multicolumn{4}{r}{\textit{Continued on next page}} \\
\endfoot
\hline \multicolumn{4}{l}{$^{*}$ 2MASS J22353504+3712131} \\
\endlastfoot
\hline    
Star & $i$ & Star & $i$\\ 
        & $Deg$ & & $Deg$ \\\hline

GJ 3942                                 &       $\textgreater$ 45  & TYC 2710-691-9 & $\sim$  90 \\
GJ 4057                                 &       $\textgreater$ 15  & 2MASS J2235$^{*}$ & $\textgreater$ 40\\
GJ 548A                 &   $\textgreater$ 50  & TYC 1795-941-1 & $\textgreater$ 40\\
GJ 3997                 &   $\textgreater$ 25  & TYC 3720-426-1 & $\textgreater$ 55\\
GJ 184                                  &       $\textgreater$ 35  & TYC 3379-1077-1 & $\textgreater$ 75\\
GJ 694.2                                &       $\textgreater$ 10   & TYC 743-1836-1 & 25 - 45 \\
GJ 9689                                 &   $\textgreater$ 25  & GJ 731 & $\textgreater$ 25 \\
GJ 740                                  &   $\textgreater$ 25  & BPM 9441 & $\textgreater$ 10 \\
GJ 685                                  &   $\textgreater$ 35  & NLTT 53166 & $\textgreater$ 40\\
GJ 720A                         &   $\textgreater$ 25  & GJ 9793 & $\textgreater$ 60 \\GJ 3822                                    &   $\textgreater$ 15  & GJ 1074 & $\textgreater$ 35 \\
GJ 606                                  &   $\textgreater$ 20  & GJ 3352 & $\textgreater$ 20 \\
GJ 412A                             &   $\textgreater$ 35  & NLTT 4188 & $\textgreater$ 30 \\
GJ 9404                                 &   $\textgreater$ 40 & SkTKM 1-650 & 15 - 55\\
TYC 2703-706-1                  &   45 - 75 &  GJ 150.1B & $\textgreater$ 20 \\
GJ 4306                                 &       $\textgreater$ 15 & GJ 3186 & $\textgreater$ 25 \\
GJ 21                                   &   $\textgreater$ 45 & GJ 272 & $\textgreater$ 25 \\
GJ 3998                                 &   $\textgreater$ 30 & GJ 686 &  $\sim$  90\\
GJ 2                                    &   $\textgreater$ 25 & GJ 4196 &  $\sim$  90\\
GJ 15A                                  &   $\textgreater$ 70 & GJ 16 & $\textgreater$ 20\\
GJ 119A                 &   $\textgreater$ 25 & GJ 3014 & $\textgreater$ 10\\
GJ 162                                  &   $\textgreater$ 50 & NLTT 10614  & $\textgreater$ 50\\
GJ 49                                   &   15 - 35  & GJ 3649 & $\textgreater$ 10 \\
GJ 156.1A                       &   $\textgreater$ 45 & GJ 895 & $\textgreater$ 20\\
GJ 9440                                 &   $\textgreater$ 25 & GJ 414B & $\textgreater$ 35\\ 
GJ 521A                 &   $\textgreater$ 20 & NLTT 52021 & $\textgreater$ 15\\
V* BRPsc                &   $\textgreater$  50 &  GJ 70 & $\textgreater$ 35\\         
GJ 625                                  &       $\sim$ 90 & GJ 2128 & $\sim$ 90\\   
GJ 47                                   &       $\textgreater$ 50 & GJ 399 & $\textgreater$ 20\\   
NLTT 21156                              &       5 - 35 & GJ 408 & $\textgreater$ 45\\
GJ 1030                 &   $\textgreater$ 15 & GJ 671 & $\sim$ 90\\
GJ 552                                  &   $\textgreater$ 20 & GJ 3117 A & $\textgreater$ 10\\
GJ 476                                  &   $\textgreater$ 40  & GJ 26 & $\textgreater$ 20 \\
&  & GJ 793 & $\textgreater$ 25\\
                        &                 & GJ 3126 & $\textgreater$ 20 \\
                        &                 & GJ 119 B & $\textgreater$ 10 \\
 \hline

\end {longtable}

\end{appendix}
\label{lastpage}

\end{document}